\documentclass[twocolumn]{aastex631} 
\usepackage{graphicx}				
\usepackage{amssymb}

\begin{document}

\title{Massive black hole formation in dense stellar environments: Enhanced X-ray detection rates in high velocity dispersion nuclear star clusters}

\author{Vivienne F. Baldassare}
\affiliation{Department of Physics and Astronomy, Washington State University, Pullman, WA 99163, USA}

\author{Nicholas C. Stone}
\affiliation{The Racah Institute of Physics, The Hebrew University of Jerusalem, Israel}

\author{Adi Foord}
\affiliation{Kavli Institute for Particle Astrophysics and Cosmology, Stanford University,
Stanford, CA 94305, USA}

\author{Elena Gallo}
\affiliation{Department of Astronomy, University of Michigan, Ann Arbor, MI 48109, USA}

\author{Jeremiah P. Ostriker}
\affiliation{Department of Astrophysical Sciences, Princeton University, Princeton, NJ 08544, USA}
\affiliation{Department of Astronomy, Columbia University, New York 10027, USA}

\correspondingauthor{Vivienne F. Baldassare} \email{vivienne.baldassare@wsu.edu}

\received{}

\begin{abstract}

We analyze \textit{Chandra X-ray Observatory} imaging of 108 galaxies hosting nuclear star clusters (NSCs) to search for signatures of massive black holes (BHs). NSCs are extremely dense stellar environments with conditions that can theoretically facilitate massive BH formation. Recent work by \cite{2017MNRAS.467.4180S} finds that sufficiently dense NSCs should be unstable to the runaway growth of a stellar mass BH into a massive BH via tidal captures. Furthermore, there is a velocity dispersion threshold ($40\;\rm{km\;s^{-1}}$) above which NSCs should inevitably form a massive BH. To provide an observational test of these theories, we measure X-ray emission from NSCs and compare to the measured velocity dispersion and tidal capture runaway timescale. We find that NSCs above the $40\;\rm{km\;s^{-1}}$ threshold are X-ray detected at roughly twice the rate of those below (after accounting for contamination from X-ray binaries). These results are consistent with a scenario in which dense, high-velocity NSCs can form massive BHs, providing a formation pathway that does not rely on conditions found only at high redshift.   

\end{abstract}


\section{Introduction}

Every galaxy with stellar mass above $\sim10^{10}\;M_{\odot}$ seems to host a massive black hole (massive BH; $M_{\rm BH} \gtrsim 10^{5}\;M_{\odot}$) at its center \citep{Kormendy:2013ve}. It remains unclear when and how these massive BHs form, although some sort of exotic high-mass seed formation mechanism may be required by the existence of the brightest high-redshift quasars \citep{2001ApJ...552..459H, 2010A&ARv..18..279V, 2020ARA&A..58...27I}. Formation channels for massive BH seeds can be divided into two categories: those that operate only at high redshift, and channels which continue to produce massive BH seeds throughout cosmic time. 

High redshift formation channels include the deaths of Population III stars \citep{2001ApJ...551L..27M, 2012ApJ...756L..19W} and direct-collapse black holes \citep{1994ApJ...432...52L, 2006MNRAS.370..289B, 2010Natur.466.1082M, 2013MNRAS.436.2989L}. Both require the low-metallicity conditions found only in the early universe. Channels in the latter category tend to invoke a gravitational runaway process which requires a dense star cluster \citep{2002ApJ...576..899P, 2004Natur.428..724P, 2004ApJ...604..632G}. Because they are dynamically driven, they can operate at any redshift. These channels are of particular interest for their promise in forming intermediate-mass BHs ($M_{\rm BH} \approx 10^{2}-10^{5}\;M_{\odot}$). Recent theoretical works have illustrated the promise of dense young star clusters for forming black holes in the pair instability mass gap from $\sim50-100\; M_{\odot}$ \citep{2020ApJ...903...45K, 2021ApJ...908L..29G}.

Nuclear star clusters (NSCs) have emerged as a promising stellar environment for facilitating the formation and/or growth of intermediate-mass BHs. NSCs reside at the centers of most galaxies with stellar masses between $\sim10^{8}-10^{10}\;M_{\odot}$ \citep{2006ApJS..165...57C, 2019ApJ...878...18S, 2021arXiv210705313H}. NSCs themselves typically have masses in the range of  $\sim10^{5}-10^{7}\;M_{\odot}$ with effective radii of a few parsecs, making them the densest known stellar environments \citep{2002AJ....123.1389B, 2004AJ....127..105B, :fz, :fs, 2012MNRAS.424.2130L, 2014MNRAS.441.3570G, 2019MNRAS.487.4285P}. 

There are two main proposed channels for the formation of NSCs: globular cluster infall due to dynamical friction \citep{1975ApJ...196..407T, Lotz:2001zr, 2008ApJ...681.1136C, Agarwal:2011fk, 2014ApJ...785...71G} and in situ star formation \citep{2006ApJ...649..692W, 2006AJ....132.2539S}. In practice, there are likely contributions from both channels, as NSCs have been found to contain multiple stellar populations \citep{2015AJ....149..170C, 2020A&ARv..28....4N, 2021arXiv210406412F, 2021arXiv210912251H}. The relative contribution of each channel likely depends on galaxy mass and environment \citep{2021arXiv210705313H}.

It was once proposed that NSCs may take the place of massive BHs as the central compact massive object in low-mass galaxies\footnote{ The absence of NSCs from the most massive galaxies is likely due to the richer merger history of the largest systems, which suffer from ``core scouring'' following major and some minor mergers \citep{1991Natur.354..212E, 1996NewA....1...35Q}.} and follow similar scaling relations with their host galaxies \citep{:uw}. However, NSCs and massive BHs have been found to coexist in some systems \citep{2008ApJ...678..116S, 2013ApJ...763...62A, 2018ApJ...858..118N, 2021arXiv211008476N}, including our own Milky Way \citep{2009A&A...502...91S, Antonini:2012uq}. The precise relationship between NSCs and massive BHs remains unclear. NSCs could form around a pre-existing BH \citep{2013ApJ...763...62A}, and/or could provide an environment in which massive BHs can more easily form. 

Several dynamical processes have been proposed for the formation of an intermediate-mass BH within a NSC \citep{2004ApJ...604..632G, Miller:2012lr, 2017MNRAS.467.4180S, 2019MNRAS.486.5008A, 2020MNRAS.498.4591F, 2021MNRAS.501.1413N, 2021arXiv210704639F}. \cite{Miller:2012lr} suggest that massive BHs will form at any epoch in a NSC with a velocity dispersion greater than a critical value of $\sigma \approx 40\; \rm{km\;s^{-1}}$. This is because above $40\; \rm{km\;s^{-1}}$, heating from primordial binaries is unable to prevent the system from undergoing core collapse. Core collapse will produce a BH subcluster; interactions within the subcluster will then either leave one or no stellar-mass BHs. If there is no BH, one will form from runaway stellar mergers \citep{2002ApJ...576..899P}. In both cases, the remaining stellar-mass BH proceeds to grow quickly through tidal capture and/or disruption events into a massive BH. 

\cite{2017MNRAS.467.4180S} expands on the above works by calculating the circumstances necessary for a NSC to grow a central stellar-mass BH into a massive one. They specifically consider a runaway tidal capture process. Tidal capture has the largest cross section for any dynamical interaction process in a dense star cluster \citep{1975MNRAS.172P..15F, 1986ApJ...310..176L}. Once there is a stellar-mass BH left at the center of the NSC, it can grow into an intermediate-mass BH through runaway tidal capture, provided that the NSC has sufficiently high central density and velocity dispersion. The runaway tidal capture will slow once the BH reaches $\sim10^{2-3}\;M_{\odot}$. At this point, BH growth will continue through tidal captures and disruptions (and possibly also standard accretion processes) until the BH reaches massive size. \cite{2017MNRAS.467.4180S} compute a timescale for this process called the tidal capture rate (TC rate, or $\rm{\dot{N}_{TC}}$). They find that many existing NSCs have rates indicating they are unstable to this runaway growth. Stated another way, given a sufficiently dense NSC, tidal capture and tidal disruption will inevitably grow a stellar-mass BH into a massive BH in less than a Hubble time.

The theoretical works described above reach a general consensus that NSCs should be able to form massive BHs through dynamical processes if they have velocity dispersions greater than $\sim40\; \rm{km\;s^{-1}}$. Indeed, there is already tentative observational evidence for this point: dynamically confirmed massive BHs are very common in galaxies with $\sigma \gtrsim 40~{\rm km~s}^{-1}$, and uncommon below this threshold (see e.g. \citealt{2017MNRAS.467.4180S} Fig. 1 or \citealt{2020ARA&A..58..257G} Fig. 3).  However, we emphasize here the circumstantial nature of existing evidence: dynamical mass measurements are challenging in the smallest galaxies, and the upper limits on BH mass produced by non-detections do not generally fall well below extrapolations of galaxy scaling relations (\citealt{2020ARA&A..58..257G}, although there are notable exceptions to this, such as M33; \citealt{2001AJ....122.2469G}).

Here, we adopt a completely different approach to explore the massive BH population amongst NSCs above and below the proposed $40~{\rm km~s}^{-1}$ threshold. Specifically, we use \textit{Chandra X-ray Observatory} (CXO) imaging to search for X-ray emission in NSCs, as sufficiently bright X-ray emission is evidence for the presence of a massive BH. We then compare the velocity dispersions and TC rates of NSCs with/without X-ray emission. This approach provides a novel consistency check for the predictions of the TC runaway theory. 

In Section 2, we describe the sample of NSCs. In Section 3, we describe the X-ray analysis and our assessment of X-ray binary contamination. Section 4 presents our results, and we discuss their implications in Section 5.


\section{Sample properties}

\begin{table*}[]
    \centering
    \begin{tabular}{c|c|c|c|c|c|c|c}
Object & RA & Dec & Dist. Modulus & Distance & $\rm{\dot{N}_{TC}}$ & Velocity Dispersion & Core Resolve \\ 
 & [deg] & [deg] & & [Mpc] & [$\rm{yr^{-1}}$] & [$\rm{km\;s^{-1}}$] & \\
\hline 
\hline
ESO138-G010 & 254.76196 & -60.21606 & 30.84 & 14.72 & 1.71E-11 & 93.39 & 0 \\ 
ESO241-G006 & 359.0625 & -43.4275 & 31.44 & 19.41 & 1.12E-09 & 23.99 & 1 \\ 
ESO359-G029 & 63.21058 & -33.00333 & 30.03 & 10.14 & 2.66E-08 & 29.47 & 1 \\ 
IC0239 & 39.11596 & 38.96903 & 30.76 & 14.19 & 4.96E-11 & 30.88 & 0 \\ 
IC0396 & 74.495792 & 68.323383 & 30.79 & 14.39 & 9.79E-12 & 100.86 & 0 \\ 
IC4710 & 277.15812 & -66.98225 & 29.75 & 8.91 & 2.37E-08 & 67.15 & 1 \\ 
IC5256 & 342.44071 & -68.69072 & 33.52 & 50.58 & 1.84E-10 & 22.01 & 1 \\ 
IC5332 & 353.6145 & -36.10156 & 29.62 & 8.39 & 2.34E-11 & 10.43 & 0 \\ 
M074 & 24.174 & 15.78333 & 29.93 & 9.68 & 1.44E-09 & 82.51 & 1 \\ 
M108 & 167.87896 & 55.67414 & 30.6 & 13.18 & 5.57E-11 & 32.07 & 0 \\ 
MCG-01-03-085 & 16.27033 & -6.21242 & 30.28 & 11.38 & 1.29E-10 & 41.97 & 0 \\ 
NGC0247 & 11.78567 & -20.76044 & 27.81 & 3.65 & 1.36E-08 & 60.35 & 1 \\ 
NGC0428 & 18.23175 & 0.98167 & 30.86 & 14.86 & 1.43E-09 & 79.84 & 0 \\ 
NGC0672 & 26.97508 & 27.43219 & 29.44 & 7.73 & 3.23E-11 & 21.04 & 0 \\ 
NGC0959 & 38.09971 & 35.49461 & 30.47 & 12.42 & 5.06E-12 & 25.49 & 0 \\ 
NGC1003 & 39.81917 & 40.87275 & 30.16 & 10.76 & 5.77E-10 & 56.43 & 1 \\ 
NGC1042 & 40.09996 & -8.43364 & 29.56 & 8.17 & 1.12E-08 & 84.03 & 1 \\ 
NGC1058 & 40.87546 & 37.34111 & 29.85 & 9.33 & 1.86E-08 & 238.72 & 1 \\ 
NGC1073 & 40.9185 & 1.37583 & 30.7 & 13.80 & 1.28E-08 & 124.05 & 0 \\ 
NGC1325A & 51.20192 & -21.33614 & 31.7 & 21.88 & 9.25E-10 & 50.19 & 1 \\ 
NGC1385 & 54.36796 & -24.50128 & 30.62 & 13.30 & 1.52E-07 & 107.57 & 1 \\ 
... & ... & ... & ... & ... & ... & ... & ... \\ 
    \end{tabular}
    \caption{Sample of 108 galaxies with NSCs and CXO observations. Distance moduli are from \cite{2016MNRAS.457.2122G}. Tidal capture rates ($\rm{\dot{N}_{TC}}$) and velocity dispersions are computed in \cite{2017MNRAS.467.4180S}. A full version of this table is available in the online version.  }
    \label{tab:props}
\end{table*}

\begin{figure}
    \centering
    \includegraphics[width=0.5\textwidth]{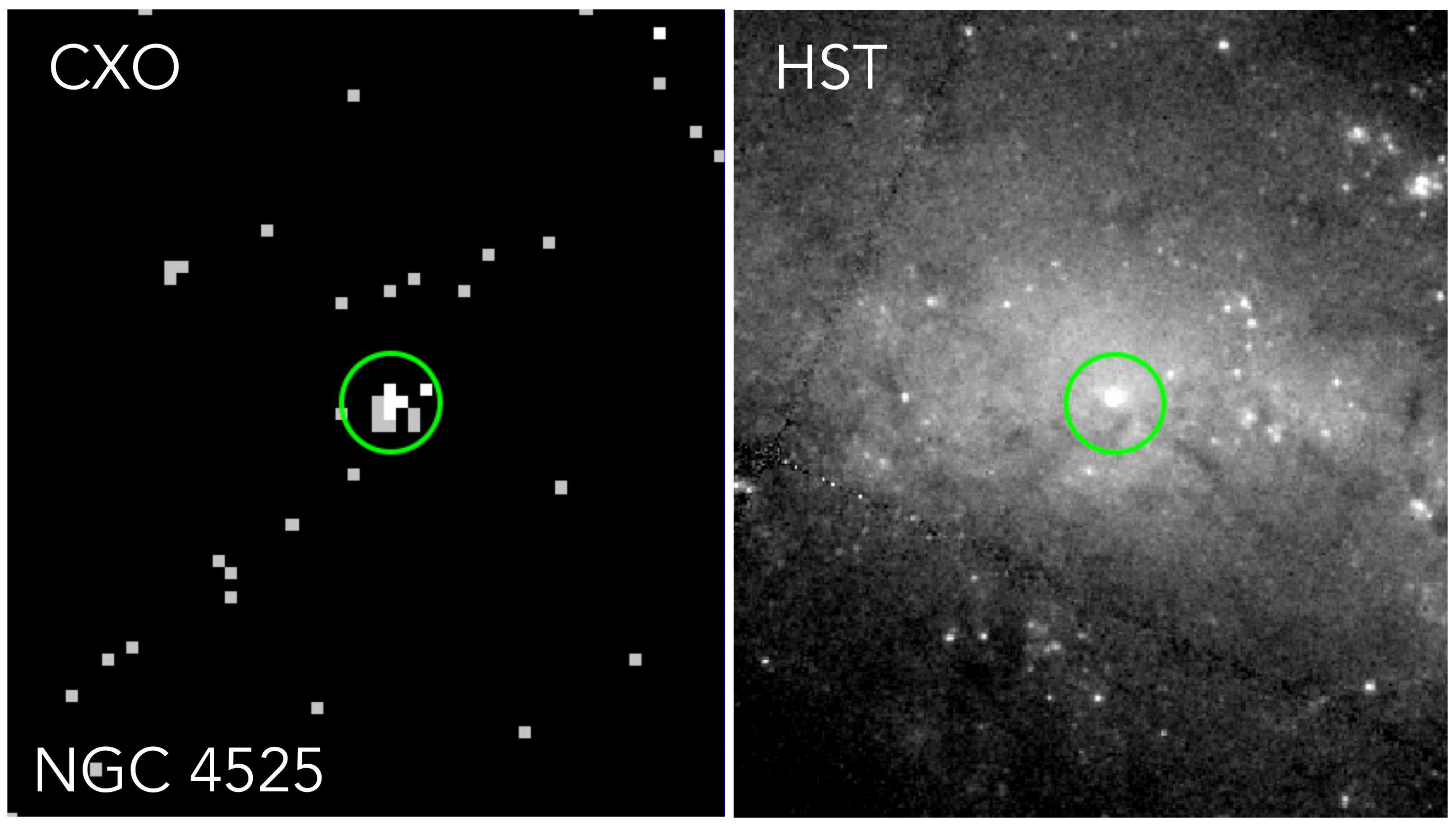}
    \caption{\textit{Chandra X-ray Observatory} imaging (left) and \textit{Hubble Space Telescope} imaging (right) for NGC 4525, a galaxy with a nuclear X-ray point source. The green circles have a radius of 2$''$ and are centered on the galaxy coordinates from \cite{2014MNRAS.441.3570G}. The HST image was taken with the WFPC2 in the F814W band. The NSC is visible in the HST image.   }
    \label{fig:ngc4525}
\end{figure}

Our parent sample contains 207 nearby ($D<50$ Mpc) galaxies hosting NSCs. All 207 have \textit{Hubble Space Telescope} observations, which are necessary in order to resolve the scales necessary to study NSCs. These are selected from the samples of \cite{2004AJ....127..105B, 2006ApJS..165...57C} and \cite{2014MNRAS.441.3570G} and were compiled by \cite{2017MNRAS.467.4180S} in order to compute tidal capture runaway timescales for actual NSCs. The parent sample consists of NSCs in both late and early-type galaxies.

All NSCs in the sample were modeled by a King profile (a tidally truncated isothermal sphere; \citealt{King:1966wd}), and have measured total masses, effective (half-light) radii, and concentration parameters. Note that for modeling, the concentration parameter (defined as $C = r_{\rm tidal}/r_{\rm core}$) was restricted to the set of $\{5, 15, 30, 100\}$ due to computational limitations. The galaxy stellar masses and NSC stellar masses for the parent sample were computed by \cite{2016MNRAS.457.2122G}. 

CXO observations were available in the archive or newly acquired for 108 out of 207 galaxies (see Section 3). These 108 galaxies are the focus of our analysis. The 108 galaxies in our sample range in distance from $2-50$ Mpc. Galaxy masses range from $10^{8} - 10^{11}\; M_{\odot}$.
The NSC masses range from $1\times10^{5}-2\times10^{8}\;M_{\odot}$, and effective radii range from $0.3-42$ pc. The sample of 108 objects is presented in Table~\ref{tab:props}.

\section{X-ray Analysis}

\begin{figure}
    \centering
    \includegraphics[width=0.5\textwidth]{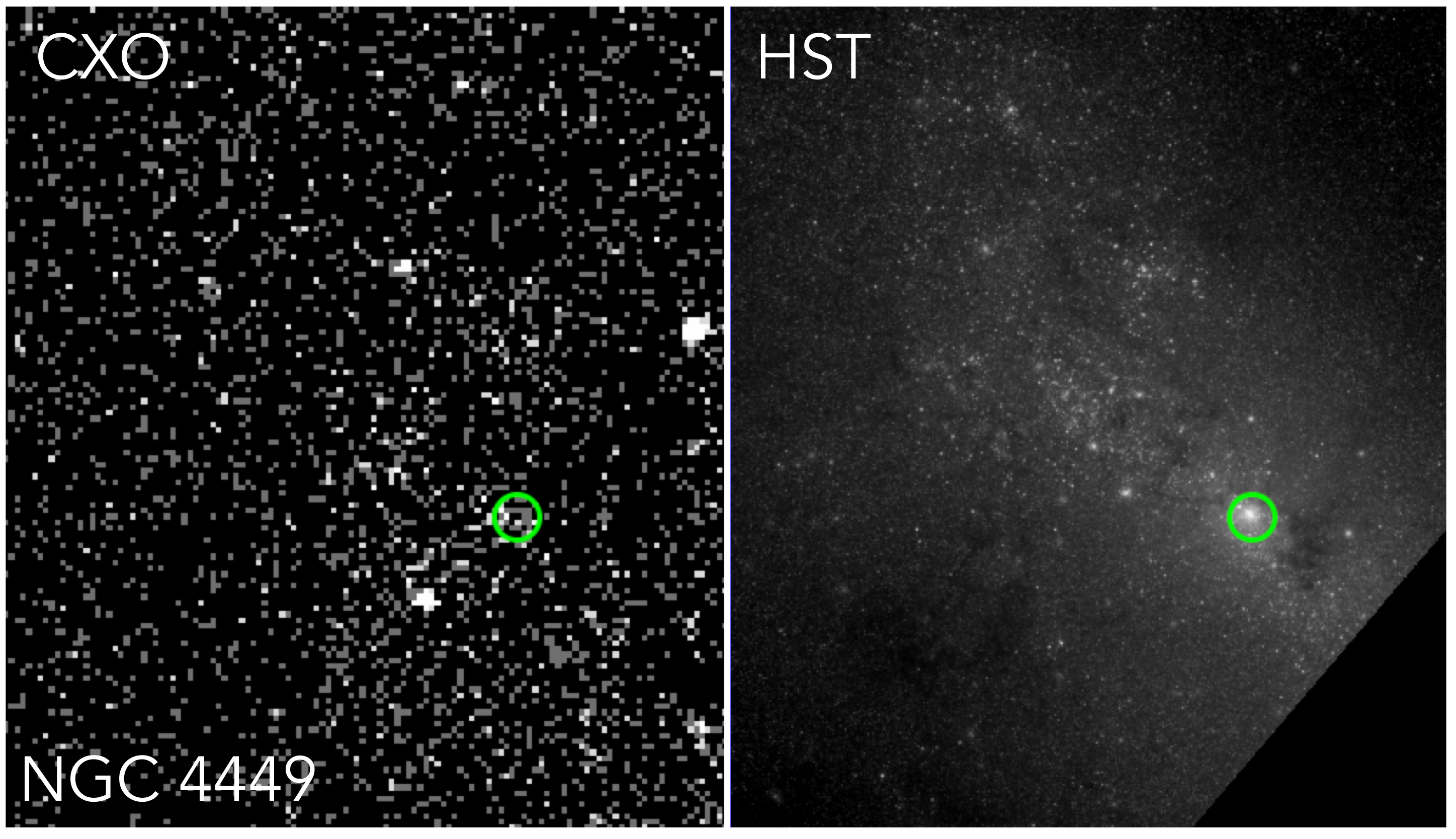}
    \caption{\textit{Chandra X-ray Observatory} imaging (left) and \textit{Hubble Space Telescope} imaging (right) for NGC 4449, a galaxy with diffuse X-ray emission in the center. The green circles have a radius of 2$''$ and are centered on the galaxy coordinates from \cite{2014MNRAS.441.3570G}. The HST image was taken with the ACS WFC in the F814W band. The NSC is visible in the HST image.   }
    \label{fig:ngc4449}
\end{figure}

We use CXO observations to study X-ray emission from our sample of NSCs. The angular resolution of CXO is necessary for determining whether any X-ray emission is coincident with the NSC. We searched the CXO archive for ACIS observations of the NSC sample. These were combined with our own program targeting the most dense, highest velocity dispersion NSCs. Fourteen objects were targeted with our program (GO 20700424; PI Baldassare), and 94 objects were in the CXO archive.

All observations were reprocessed and analyzed with the Chandra Interactive Analysis of Observations software (CIAO; version
5.13). We generate an initial source list using CIAO WAVDETECT. We then correct the astrometry by cross-matching the X-ray source list with sources in the USNO B-1 catalog. Matches were required to be within $2''$ of one another, and we required three or more matches to apply the astrometry correction. We next filtered out any background flares in the observations. We then applied an energy filter in the 0.5-7 keV range and reran WAVDETECT with a threshold significance of $10^{-6}$, which corresponds to one false detection over a single ACIS chip. 

Using SRCFLUX, we computed count rates, fluxes, and uncertainties in the soft (0.5-1.2 keV), broad (0.5-7 keV), and hard (2-10 keV) bands. We take the coordinates given by \cite{2014MNRAS.441.3570G} to be the NSC position. If there was a WAVDETECT source coincident with the galaxy center, we extracted the counts in a circular region with a radius of 2" at the WAVDETECT source position. If there was not a source detected at the nucleus, we extracted counts in the same size circular region centered on the \cite{2014MNRAS.441.3570G} coordinates. For the background region, we use a source-free annulus with inner and outer radii of $\sim20$ and $\sim35$ arcseconds, respectively. We compute the unabsorbed model flux for a power law spectrum with $\Gamma=1.8$ and the Galactic $n_{\rm H}$ value returned by the CIAO colden tool. 

In all, 46/109 objects are X-ray detected. However, 5 of the 46 have diffuse X-ray emission rather than point sources. We do not consider these five as possible BHs and do not analyze them with the other 41 sources. An example of a galaxy with a nuclear X-ray point source is shown in Figure~\ref{fig:ngc4525} and an example of a diffuse source is shown in Figure~\ref{fig:ngc4449}. 

Of the 41 NSCs with X-ray point sources, 35 are detected in both the 0.5-7 and 2-10 keV bands, and 6 are only significantly detected in the 0.5-7 keV band. Luminosities in the 0.5-7 keV band range from to $2.0\times10^{37} - 2.0\times10^{42}\; \rm{erg\;s^{-1}}$, with a median luminosity of $1.2\times10^{39} \rm{erg\;s^{-1}}$. The X-ray luminosities and upper limits are given in Table~\ref{tab:xray}.

\subsection{X-ray binary contamination}

Given the X-ray luminosities found for the NSCs in our sample, we must consider the possibility of contamination from X-ray binaries. While it is difficult to state definitively whether any particular source is a massive BH or an X-ray binary, we can estimate the likely X-ray luminosity from low-mass X-ray binaries (LMXBs) and high-mass X-ray binaries (HMXBs) for each NSC. The contribution from LMXBs traces the cumulative star formation history of the galaxy, and is thus proportional to the total stellar mass \citep{:ut}. The contribution from HMXBs traces recent star formation in the galaxy, and is proportional to the galaxy's star formation rate \citep{2003MNRAS.339..793G, 2012MNRAS.419.2095M}. 

\cite{2019ApJS..243....3L} uses a sample of 38 nearby galaxies to constrain the scaling relations for X-ray binaries. Their sample spans a wide range in galaxy morphology, stellar mass, star formation rate, and metallicity. It includes both nucleated and non-nucleated galaxies; several of the galaxies in their sample overlap with our sample of 108 galaxies. The expected X-ray luminosity from LMXBs and HMXBs is given by 
\begin{equation}
L_{\rm X} = \alpha M_{\ast} + \beta\rm{SFR}.    
\end{equation}
The best fit values from \cite{2019ApJS..243....3L} are $\alpha = 1.8\times10^{29} \rm{erg\;s^{-1}\;M_{\odot}^{-1}}$ and $\beta = 5.1\times10^{39} \rm{erg\;s^{-1}\;(M_{\odot}\;yr^{-1})^{-1}}$. We compute the expected X-ray binary luminosity for each of our galaxies. Note that this uses the 0.5-8 keV luminosity; we compute the 0.5-8 keV luminosities for our X-ray sources in order to compare to the estimated luminosities from \cite{2019ApJS..243....3L}.  

Galaxy stellar masses are taken from \cite{2014MNRAS.441.3570G}. To compute the SFRs, we use the formalism from \cite{2012ARAA..50..531K}, where 

\begin{equation}
    \log \dot{M}_{\ast}(M_{\odot}\;yr^{-1}) = \log L_{\rm FUV, corr} - 43.35. 
\end{equation}

The FUV luminosity is corrected for dust extinction using the 25$\micron$ luminosity, where $L_{\rm FUV, corr} = L_{\rm FUV, obs} + 3.89\times L_{25\micron}$.

We use GALEX data \citep{2005ApJ...619L...1M} to measure the FUV luminosities for each galaxy in our sample. We measure the FUV flux within an aperture with a radius equal to the galaxy radius reported in \cite{2014MNRAS.441.3570G} (their $R_{25}$, from HyperLeda; \citealt{2003A&A...412...45P}). We use W4 magnitudes from AllWISE \citep{2010AJ....140.1868W} to compute the $25\micron$ flux density and correct the FUV luminosity. 

We are interested in the expected X-ray binary luminosity within the CXO point spread function. Following similar analyses presented in \cite{2017ApJ...841...51F} and \cite{2019ApJ...874...77L}, we assume that the galaxy mass traces the light profile, and measure the fraction of galaxy light in the central 2$''$ for each system using available HST imaging. We note that galaxies were imaged with different cameras and filters. The median fraction of light in the central 2$''$ is 0.035. For a handful of cases, the galaxy is larger than the HST image field of view; for these galaxies, we take the fraction to be 0.05, slightly more conservative than the median value.

Putting the above steps together, we find expected X-ray binary luminosities in the central 2$''$ ranging from $6\times10^{36}-3\times10^{39}\rm{erg\;s^{-1}}$. The ratio of observed 0.5-8 keV luminosity to that expected from X-ray binaries ranges from $0.09$ to $3980$, with a median ratio of 6.27. Table~\ref{tab:xrbs} gives the expected X-ray binary luminosities and ratios of observed-to-expected luminosity.

A caveat to this analysis is that it does not take into account additional potential emission due to the NSC environment, i.e., preferential LMXB/HMXB formation in the central parsecs. Additionally, a possible formation channel for NSCs is through inspirals of globular clusters, where X-ray binary formation is efficient \citep{1975ApJ...199L.143C, :yf}. As our formation channel is driven by runaway tidal capture LMXB formation in NSCs, this may seem contradictory. However, we consider as a comparison case the central pc of the Milky Way galaxy, which indeed possesses an overabundance of LMXBs \citep{2018Natur.556...70H}, that may themselves be BH-LMXBs that have formed through tidal capture \citep{2018MNRAS.478.4030G}. Except during brief periods of LMXB outburst \citep{2005ApJ...622L.113M}, the combined X-ray luminosity of this LMXB subcluster is $\lesssim 10^{34} {\rm erg~s}^{-1}$ , likely due to the low mass-transfer rates in most tidal capture LMXBs. Furthermore, the Milky Way NSC lacks clear evidence for an overabundance of NS-LMXBs, which form at elevated rates in globular clusters due to chaotic binary-single scatterings \citep{2008MNRAS.386..553I}. This signifies (i) a low rate of NS-LMXB formation in NSCs relative to globular clusters, likely due to lower binary fractions resulting from the higher velocity dispersion environment of the galactic nucleus\footnote{However, the Milky Way NSC has a high velocity dispersion and low binary fraction at least in part because of the presence of a SMBH; in an NSC lacking a central massive BH, the binary fraction might be high enough to favor dynamical NS-LMXB formation in three-body scatterings.}, and (ii) that whatever NS-LMXBs were brought in through globular inspiral have since deactivated. As the typical NS-LMXB lifetime is $\sim 1$ Gyr \citep{2008MNRAS.386..553I}, the latter conclusion would be unsurprising if most globular inspiral events happened in the distant past.

Given the generally low X-ray luminosity of LMXBs in the Milky Way NSC and lacking a NSC-specific XRB luminosity estimator, we proceed with our estimates of probable X-ray emission based on SFR and stellar mass within the PSF. We note, however, that we may be under-predicting the true total XRB luminosity of some NSCs if their dynamics deviates strongly from that of the Milky Way NSC.

Going forward, we consider anything with an observed luminosity a factor of 2 higher than expected to be due to a massive BH (e.g., \citealt{2020arXiv200103135B}). Objects with observed luminosities less than twice the expected are treated as non-detections. 
This removes 10 detections from the $\sigma>40\rm{km\;s^{-1}}$ regime, and 2 from the $\sigma<40\rm{km\;s^{-1}}$ regime. These tend to be lower luminosity systems; the median 0.5-7 keV luminosity for the likely X-ray binary systems is $\sim10^{38}\rm{erg\;s^{-1}}$, compared to $10^{39} \rm{erg\;s^{-1}}$ for the more secure massive BHs. Our results do not change if we use a more conservative factor of 3 instead of 2. 

\begin{rotatetable*}
\begin{deluxetable*}{cccccccccc}
    \tablecolumns{10}
    \tablewidth{0pt}
    \tablecaption{X-ray properties \label{tab:xray}}
    \tablehead{\colhead{Object} & \colhead{ObsID} & \colhead{Exp Time [ks]} & \colhead{$L_{\rm{0.5-7}}$} & \colhead{$L_{\rm{0.5-7, lower}}$} & \colhead{$L_{\rm{0.5-7, upper}}$} & \colhead{$L_{\rm{2-10}}$} & \colhead{$L_{\rm{2-10, lower}}$} & \colhead{$L_{\rm{2-10, upper}}$} & \colhead{Diffuse}  }
\startdata
ESO138-G010 & 14800 & 9.84 & 38.04 & 37.58 & 38.37 & - & - & 38.36 & N  \\ 
ESO241-G006 & 17004 & 4.7 & - & - & 38.32 & - & - & 38.71 & N  \\ 
ESO359-G029 & 21468 & 3.07 & - & - & 38.06 & - & - & 38.34 & N  \\ 
IC0239 & 7131 & 4.53 & - & - & 38.01 & - & - & 38.47 & N \\ 
IC0396 & 7134 & 4.87 & 39.08 & 38.94 & 39.22 & 38.69 & 38.24 & 39.02 & N  \\ 
IC4710 & 9877 & 15.25 & - & - & 37.28 & - & - & 37.52 & N  \\ 
IC5256 & 17001 & 1.65 & - & - & 39.58 & - & - & 39.99 & N  \\ 
IC5332 & 2067 & 55.24 & - & - & 36.47 & - & - & 37.02 & N  \\ 
M074 & 4753 & 5.28 & 38.05 & 37.71 & 38.32 & 38.04 & 37.28 & 38.49 & N  \\ 
M108 & 2025 & 59.36 & 37.61 & 37.42 & 37.78 & 37.41 & 36.44 & 37.80 & Y  \\ 
MCG-01-03-085 & 12981 & 9.82 & - & - & 37.48 & - & - & 38.11 & N  \\ 
NGC0247 & 17547 & 5.01 & 39.19 & 39.16 & 39.22 & 39.17 & 39.12 & 39.22 & N  \\ 
NGC0428 & 16978 & 3.99 & - & - & 38.17 & - & - & 38.55 & N  \\ 
NGC0672 & 7090 & 2.15 & - & - & 40.90 & - & - & 41.46 & N  \\ 
NGC0959 & 7111 & 2.18 & - & - & 38.17 & - & - & 38.65 & N  \\ 
NGC1003 & 7116 & 2.67 & - & - & 38.03 & - & - & 38.50 & N \\ 
NGC1042 & 12988 & 29.01 & 38.09 & 37.98 & 38.18 & 38.16 & 37.99 & 38.31 & N  \\ 
NGC1058 & 387 & 2.41 & 38.00 & 37.45 & 38.37 & 38.34 & 37.57 & 38.80 & N \\ 
NGC1073 & 4686 & 5.74 & 38.90 & 38.75 & 39.03 & 39.03 & 38.80 & 39.22 & N  \\ 
NGC1325A & 7841 & 5.09 & - & - & 38.28 & - & - & 38.76 & N \\ 
NGC1385 & 21473 & 5.04 & 42.22 & 42.09 & 42.32 & 42.28 & 42.11 & 42.40 & N  \\ 
NGC1483 & 16981 & 2.51 & - & - & 39.01 & - & - & 39.37 & N \\ 
NGC1487 & 21469 & 2.05 & - & - & 38.12 & - & - & 38.40 & N  \\ 
NGC1493 & 7145 & 10.03 & 38.76 & 38.64 & 38.85 & 38.47 & 38.18 & 38.70 & N  \\ 
... & ... & ... & ... & ... & ... & ... & ... & ... & ... \\ 
\enddata
\tablecomments{Columns 4-6 give the log of the 0.5-7 keV luminosity (in $\rm{erg\;s^{-1}})$), lower limit, and upper limit, respectively. Columns 7-9 give the log of the 2-10 keV luminosity (in $\rm{erg\;s^{-1}})$), lower limit, and upper limit, respectively. Column 10 states whether the detected X-ray emission was diffuse. The complete table is available in the online version of this paper. }
\end{deluxetable*}
\end{rotatetable*}

\begin{figure*}
    \centering
    \includegraphics[width=0.9\textwidth]{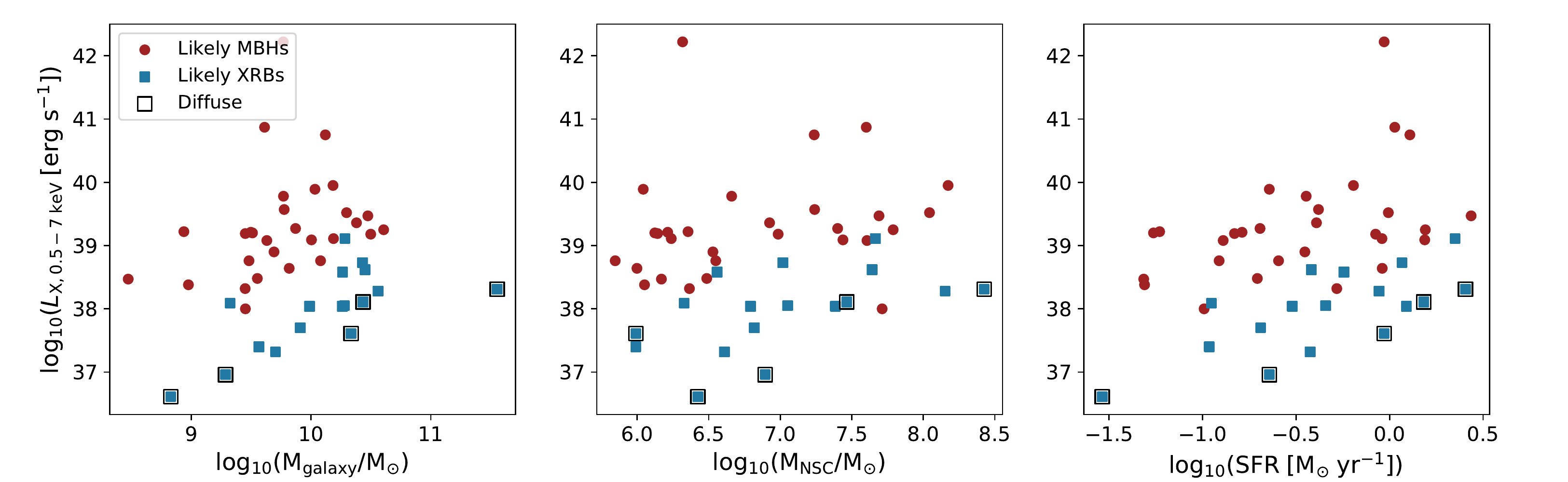}
    \caption{X-ray luminosity versus galaxy properties for X-ray detected galaxies that are likely massive BHs (red circles), and likely XRBs (blue squares). From left to right, we show galaxy stellar mass, NSC stellar mass, and galaxy star formation rate. At a given X-ray luminosity, the likely XRBs are in more massive galaxies/NSCs and have higher star formation rates. We also emphasize the galaxies with diffuse X-ray emission. These are all likely XRB systems and have some of the lowest detected X-ray luminosities.}
    \label{fig:xrb_pan}
\end{figure*}

In Figure~\ref{fig:xrb_pan}, we show the X-ray luminosity versus galaxy mass, NSC mass, and star formation rate for likely massive BHs and likely XRBs. At a given X-ray luminosity, the likely XRBs tend to be in more massive galaxies and those with higher SFRs. This is expected as more massive galaxies will have more stellar mass within the central 2$''$.

In addition to the 12 point source objects we treat as likely X-ray binaries, all five of the galaxies with diffuse X-ray emission have X-ray luminosities less than the expected luminosity from X-ray binaries. Their luminosities range from $\sim0.1-0.5$ times the expected X-ray binary luminosity. 

\begin{table*}[]
    \centering
    \begin{tabular}{c|c|c|c|c|c}
Object & $M_{\ast,\rm{galaxy} }$ & SFR & $L_{\rm{XRBs, 0.5-8 keV}}$ & $L_{\rm{source, 0.5-8 keV}}$ & $L_{\rm{source}}$ / $L_{\rm{XRBs}}$ \\ 
 & [$M_{\odot}$] & [$M_{\odot}\;\rm{yr^{-1}}$] & $[\log_{10}(\rm{erg\;s^{-1}})]$ & $[\log_{10}(\rm{erg\;s^{-1}})]$ & \\
\hline 
\hline
ESO138-G010 & 1.82E+10 & 0.3018 & 38.35 & 38.11 & 0.58 \\ 
IC0396 & 4.23E+09 & 0.1290 & 38.18 & 39.15 & 9.33 \\ 
M074 & 1.90E+10 & 0.4551 & 38.46 & 38.12 & 0.46 \\ 
M108 & 2.16E+10 & 0.9364 & 38.64 & 37.68 & 0.11 \\ 
NGC0247 & 2.82E+09 & 0.1482 & 36.79 & 39.26 & 296.59 \\ 
NGC1042 & 2.11E+09 & 0.1115 & 37.90 & 38.16 & 1.84 \\ 
NGC1058 & 2.83E+09 & 0.1020 & 37.63 & 38.07 & 2.74 \\ 
NGC1073 & 4.90E+09 & 0.3525 & 38.05 & 38.97 & 8.34 \\ 
NGC1385 & 5.85E+09 & 0.9353 & 38.69 & 42.29 & 3981.04 \\ 
NGC1493 & 3.02E+09 & 0.2550 & 37.97 & 38.82 & 7.23 \\ 
NGC1559 & 5.88E+09 & 0.3583 & 38.16 & 39.85 & 49.28 \\ 
NGC1566 & 1.52E+10 & 0.6410 & 38.48 & 40.02 & 35.15 \\ 
NGC2139 & 1.01E+10 & 1.5415 & 38.22 & 39.16 & 8.60 \\ 
NGC2207 & 1.98E+10 & 0.9849 & 38.63 & 39.59 & 9.10 \\ 
NGC2748 & 1.54E+10 & 0.9105 & 38.57 & 39.18 & 4.11 \\ 
NGC2835 & 5.05E+09 & 0.3765 & 37.93 & 37.39 & 0.29 \\ 
... & ... & ... & ... & ... & ... \\ 
    \end{tabular}
    \caption{X-ray binary contamination assessment. Column 4 gives the expected 0.5-8 keV luminosity from XRBs following the scalings derived in \cite{2019ApJS..243....3L}. Column 5 gives the 0.5-8 keV luminosity of the detected X-ray source. Column 6 gives the ratio of the source luminosity to the expected XRB luminosity. The complete table will be available in the online version of the paper. }
    \label{tab:xrbs}
\end{table*}

\subsection{Hardness ratios}

\begin{figure}
\centering
    \includegraphics[width=0.45\textwidth]{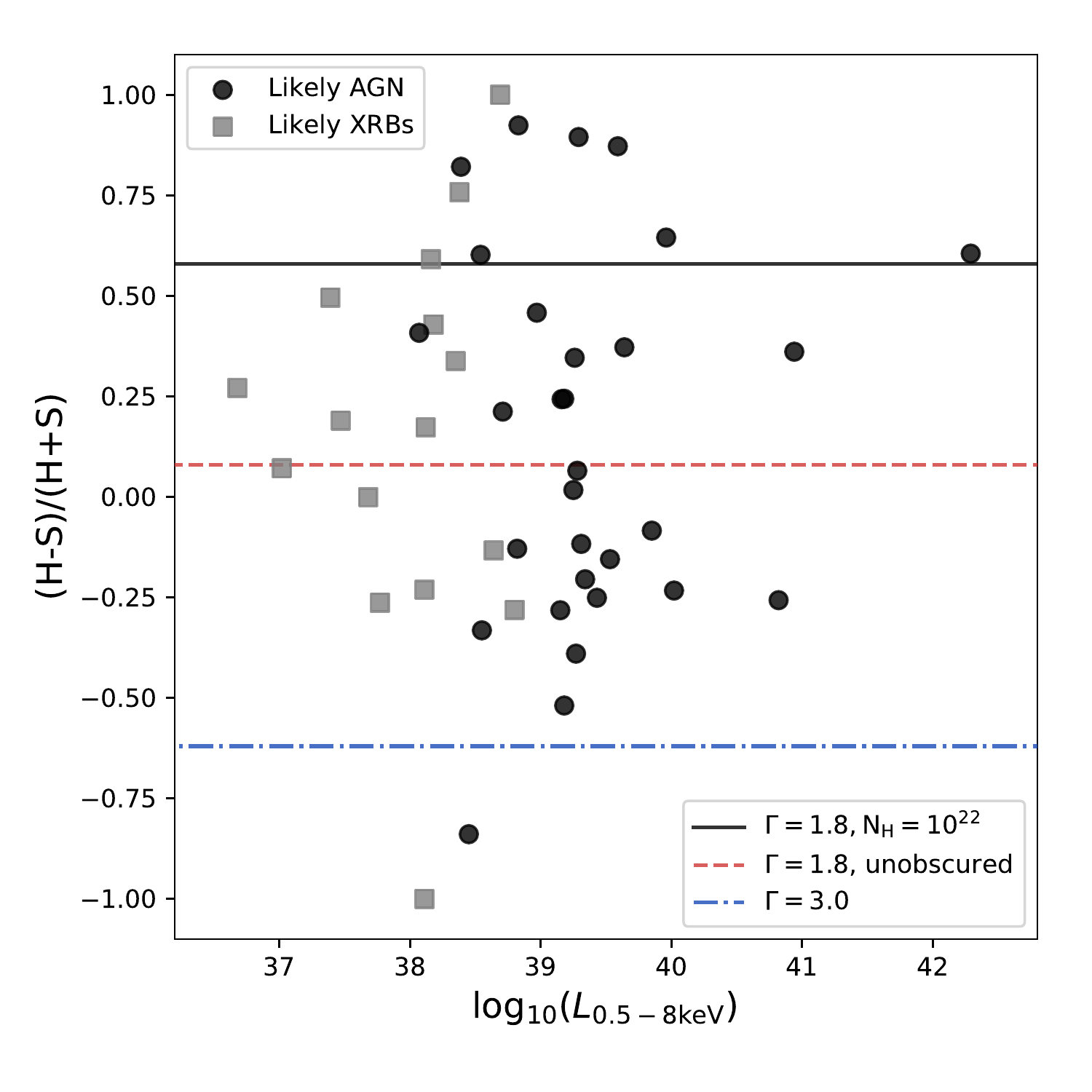}
    \caption{Hardness ratio versus 0.5-8 keV X-ray luminosity. H is the flux in the 2-7 keV band, and S is the flux in the 0.5-2 keV band. We use WebPIMMs to compute hardness ratios for sources with various power law indices. A power law index of $\Gamma = 1.8$ is typical of AGN; we show the hardness ratios for this value for an obscured and unobscured source. A power law index of $\Gamma = 3$ reflects a softer X-ray spectrum consistent with high Eddington fraction X-ray binaries \citep{2015MNRAS.447.1692Y}. }
    \label{fig:HR_v_Lx}
\end{figure}

We compute hardness ratios for our sources using the aperture fluxes reported by SRCFLUX. The hardness ratio is given by $[F(H) - F(S)]/[F(H) + F(S)]$ where $F(H)$ is the flux in a hard band and $F(S)$ is the flux in a soft band. Here, we use the 0.5-2 keV band as the soft band and the 2-7 keV band as the hard band. In Figure~\ref{fig:HR_v_Lx}, we show the hardness ratios for the likely active galactic nuclei (AGN) and likely XRBs. The samples of likely AGN and XRBs both span a wide range in hardness ratios (from -1 to 1), with similar distributions. Many are consistent with a typical AGN spectrum with $\Gamma = 1.8$.

\begin{figure*}
    \centering
    \includegraphics[width=\textwidth]{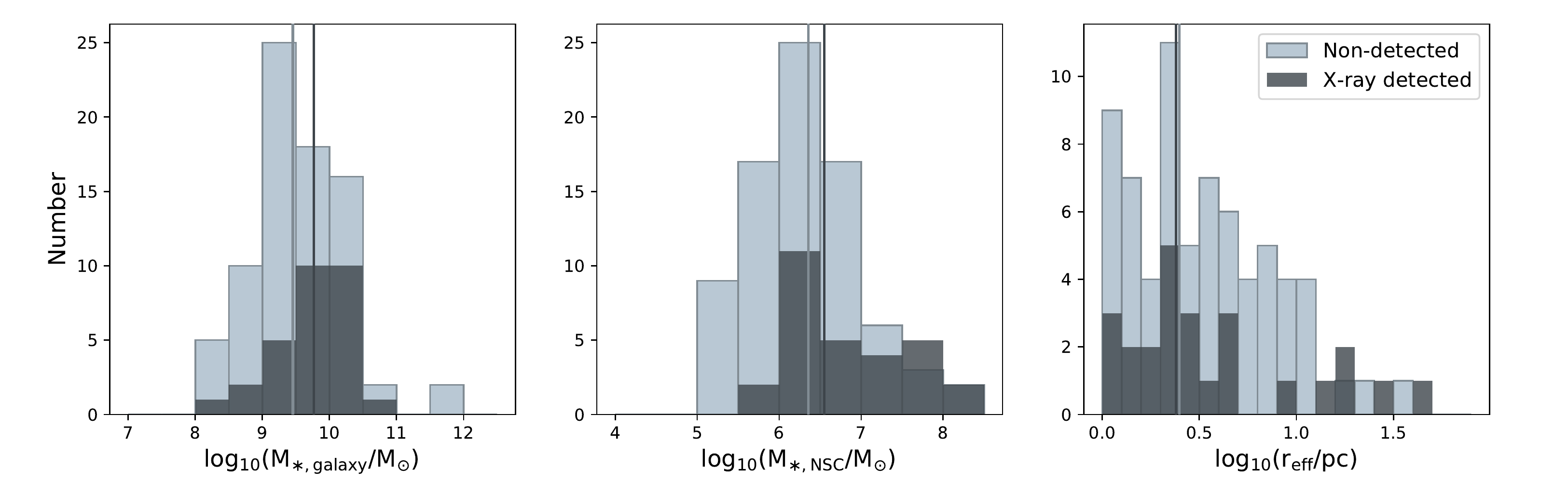}
    \caption{Properties of galaxies with and without X-ray detected NSCs. From left to right, we show galaxy stellar mass, NSC stellar mass, and NSC effective radius. The solid lines show the median values of each distribution. }
    \label{fig:hist}
\end{figure*}

\section{Results}

As described in Section 3, 41 NSCs have X-ray point sources coincident with the NSC position. 12 of these have X-ray luminosities close to the expected values for X-ray binary emission based on the scaling relations derived by \cite{2019ApJS..243....3L}. In this section, we explore the properties, velocities dispersions, and tidal capture rates of NSCs with/without X-ray emission, treating the 12 likely X-ray binary contaminants as non-detections. Henceforth, ``X-ray detected" refers to galaxies with X-ray detections likely to be from a massive BH.

\subsection{Galaxy properties} 

In Figure~\ref{fig:hist}, we show the distributions of galaxy mass, NSC mass, and effective radius for galaxies with and without X-ray detected NSCs. This is important for understanding what factors influence a NSC hosting an X-ray active BH. The masses and effective radii were measured by \cite{2016MNRAS.457.2122G} based on multi-band HST imaging.

The median galaxy stellar masses for the X-ray detected and non-detected galaxies are $5.9\times10^{9}\;M_{\odot}$ and $2.8\times10^{9}\;M_{\odot}$, respectively. The X-ray detected NSCs tend towards slightly higher-mass galaxies. The median NSC stellar masses for the X-ray detected and non-detected galaxies are $3.5\times10^{6}\;M_{\odot}$ and $2.3\times10^{6}\;M_{\odot}$, respectively, and the distribution for X-ray detected NSCs skews towards slightly higher NSC masses. Finally, the median effective radii for the two samples are similar; 2.4 and 2.5 pc.

\begin{figure}
    \centering
    \includegraphics[width=0.5\textwidth]{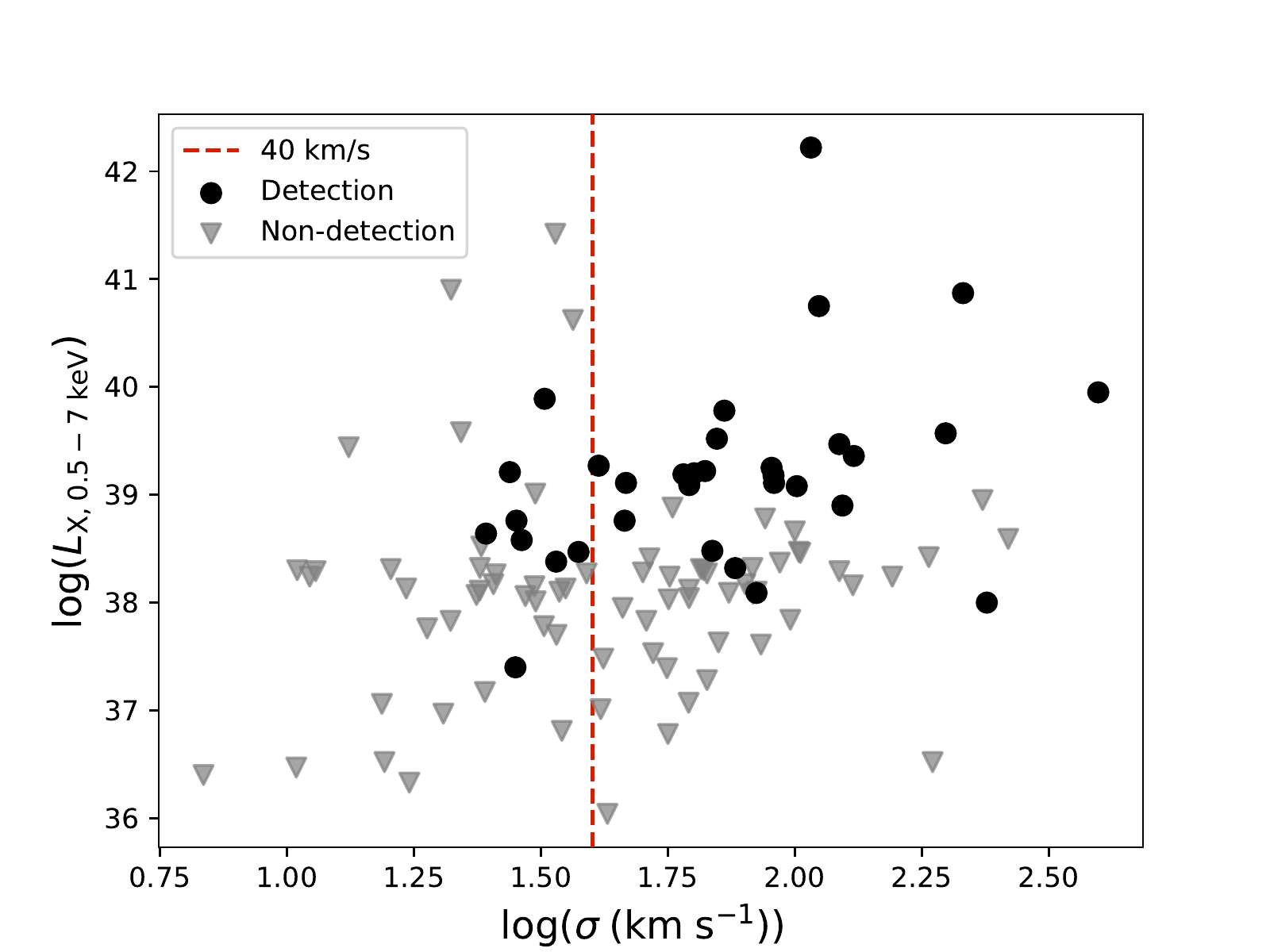}
    \caption{X-ray luminosity (0.5-7 keV) versus NSC 1-D velocity dispersion. Detections are shown as black circles, and upper limits are represented by gray triangles. The vertical dashed red line marks the $40\;\rm{km\;s^{-1}}$ threshold. }
    \label{fig:lx_disp}
\end{figure}

\subsection{Detection rate versus velocity dispersion}

\begin{figure*}
    \centering
    \includegraphics[width=\textwidth]{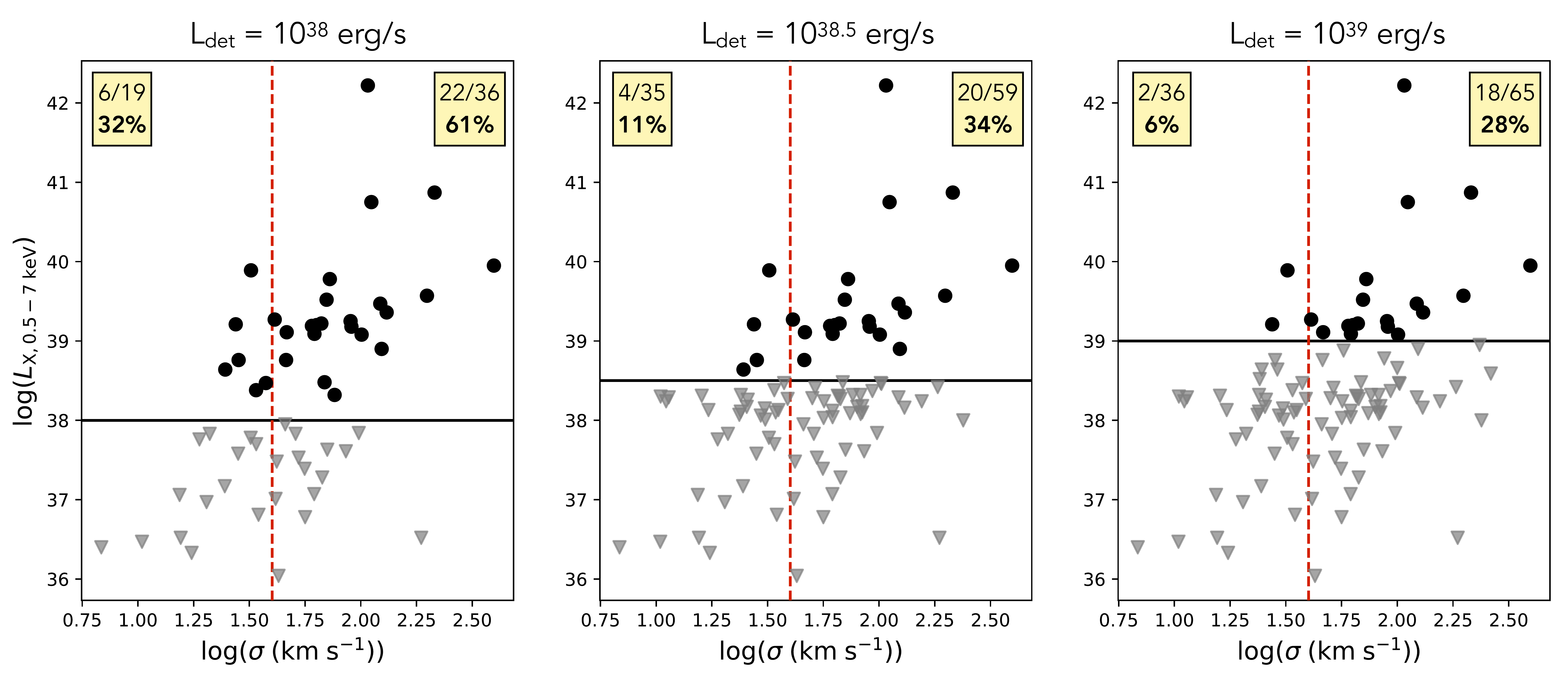}
    \caption{Same as Figure~\ref{fig:lx_disp}, but with different imposed detection limits, $L_{\rm det}$. The horizontal black solid line shows $L_{\rm det}$ for each panel. Any X-ray detection below that limit is treated as an upper limit. We compute the detection fraction for NSCs with 1-D velocity dispersions above and below $40\;\rm{km\;s^{-1}}$ for each value of $L_{\rm det}$. The detection fractions for each are shown at the top of each panel. The detection fractions for NSCs in the high velocity dispersion regime are consistently and significantly higher. We note that galaxies with X-ray emission likely to be due to X-ray binaries are also treated as non-detections, since they do not have X-ray detections consistent with the presence of a massive BH. }
    \label{fig:panels}
\end{figure*}

Our goal in this section is to compare the X-ray properties of NSCs with velocity dispersions above and below the $40\;\rm{km\;s^{-1}}$ threshold. The cluster averaged velocity dispersion $\sigma$ is defined by $\sigma=\sqrt{G\;M_{\rm tot} / (3\; r_{\rm eff})}$ where $r_{\rm eff}$ is the half-mass radius from the best-fit King model. In Figure~\ref{fig:lx_disp}, we show the X-ray luminosity versus velocity dispersion for the entire sample. 

\begin{figure}
    \centering
    \includegraphics[width=0.5\textwidth]{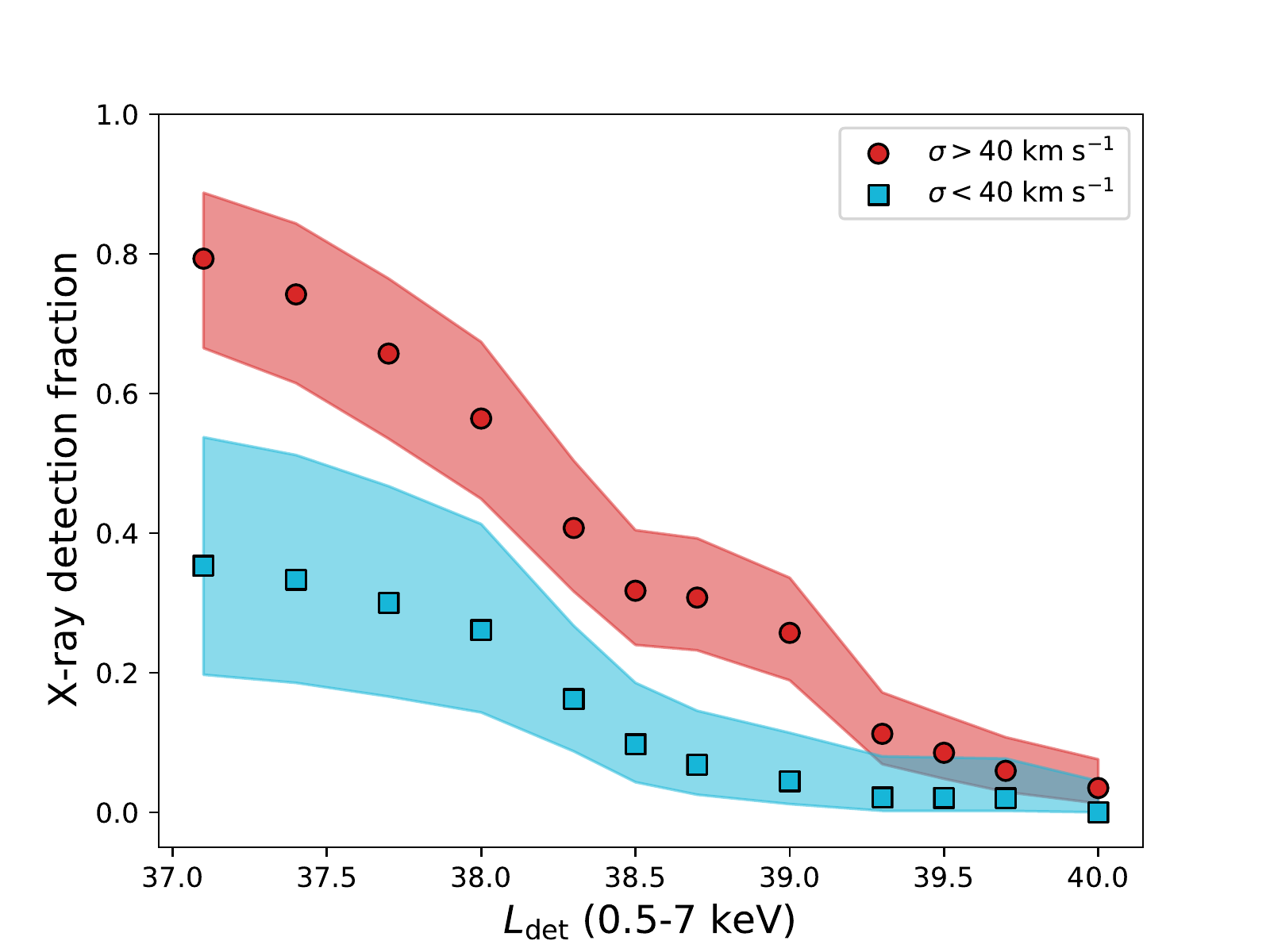}
    \caption{X-ray detection fraction versus $L_{\rm det}$ for NSCs with velocity dispersions above (red circles) and below (blue squares) $40\rm{km\;s^{-1}}$. This is the nominal threshold above which NSCs are unstable to core collapse and will form a massive BH. The shaded region reflects the 90\% confidence limits \citep{Gehrels:1986kx}.}
    \label{frac_v_limit}
\end{figure}

However, we cannot take the detection fractions above and below this velocity dispersion threshold at face value. The X-ray sample is non-homogeneous and observations have different limiting fluxes. To account for this, we compute the fraction for samples with artificially imposed X-ray detection limits. For each imposed detection limit $L_{\rm det}$, we exclude any non-detections with upper limits greater than that value (since we cannot say whether they are detected down to a limit of $L_{\rm det}$). We count the number of NSCs with measured X-ray luminosities greater than $L_{\rm det}$, and treat anything with an X-ray luminosity lower than $L_{\rm det}$ as a non-detection. Note that any galaxies with X-ray luminosities likely to be due to X-ray binaries are also treated as non-detections. We then compare the number of detected NSCs against those with upper-limits lower than $L_{\rm det}$. We repeat this for different values of $L_{\rm det}$. Figure~\ref{fig:panels} demonstrates how the sample looks for different values of $L_{\rm det}$. Figure~\ref{frac_v_limit} shows X-ray detection fraction versus $L_{\rm det}$ for NSCs above and below the velocity dispersion threshold. 

From Figures~\ref{fig:panels} and \ref{frac_v_limit}, we can see that for all values of $L_{\rm det}$, the detection fraction is significantly higher for the $\sigma>40\rm{km\;s^{-1}}$ sample.  At $L_{\rm det} = 10^{39} \rm{erg\;s^{-1}}$, the detection fraction for $\sigma<40\rm{km\;s^{-1}}$ is $6^{+8}_{-5}\%$ and the detection fraction for $\sigma>40\rm{km\;s^{-1}}$ is $28\% \pm 8\%$ (uncertainties are given for the 90\% confidence limit).

As expected, the detection fraction increases as we increase our sensitivity (i.e., lower the value of $L_{\rm det}$), but does so similarly for the low and high velocity dispersion regimes.

\subsection{Detection rate versus tidal encounter rate}

We also look at the X-ray detection rate versus the tidal capture (TC) rates determined by \cite{2017MNRAS.467.4180S}. The tidal capture rate $\rm{\dot{N}_{TC}}$ is given by equation 14 in \cite{2017MNRAS.467.4180S} and depends on the mass of the initial stellar-mass black hole, the velocity dispersion, central cluster density, and average mass and radius of the cluster stars. Some of the NSCs in our sample have only lower limits on the TC rate due to the discrete concentration values used in fitting the light profiles of the NSCs (objects with ``Core Resolve" $= 0$ in Table~\ref{tab:props}). We exclude these from computations of the detection rate since we cannot say conclusively whether they are unstable to runaway TC over a Hubble time. We also note that the TC rates are uncertain by $\sim$ one order of magnitude for this reason. We refer the reader to Section 6.1 of \cite{2017MNRAS.467.4180S} for a detailed discussion of these uncertainties. 

\begin{figure}
    \centering
    \includegraphics[width=0.5\textwidth]{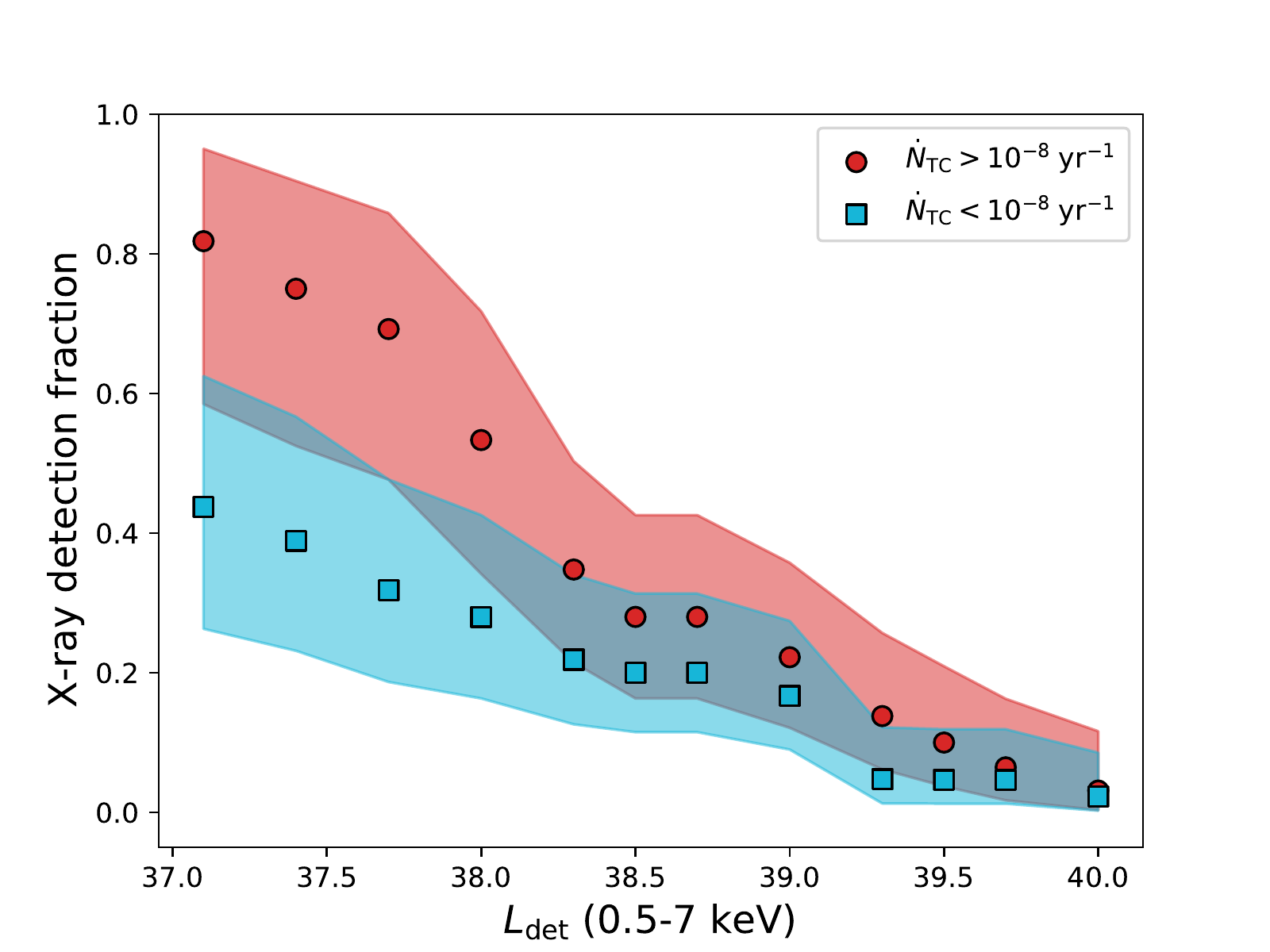}
    \caption{X-ray detection fraction versus $L_{\rm det}$ for NSCs with velocity dispersions above (red circles) and below (blue squares) $\dot{N}_{\rm TC}=10^{-8}\; \rm{yr^{-1}}$. This is the nominal threshold above which a stellar-mass BH in a NSC is unstable to growth through the tidal capture runaway \citep{2017MNRAS.467.4180S}. The shaded region reflects the 90\% confidence limits \citep{Gehrels:1986kx}.}
    \label{frac_v_limit_ntc}
\end{figure}

In Figure~\ref{frac_v_limit_ntc}, we plot the X-ray detection fraction versus $L_{\rm det}$ for NSCs with TC rates above and below $\dot{N}_{\rm TC}=10^{-8}\; \rm{yr^{-1}}$. NSCs with TC rates $\dot{N}_{\rm TC}>10^{-8}\; \rm{yr^{-1}}$ should be unstable to runaway growth of a stellar-mass BH through TC. While the fractions are higher for the $\dot{N}_{\rm TC}>10^{-8}\; \rm{yr^{-1}}$ sample, the uncertainties are larger due to the smaller number of objects in the sample. The two groups are consistent with one another within the 90\% confidence intervals. In order to confirm whether the X-ray detection fractions are truly higher amongst NSCs with $\dot{N}_{\rm TC}>10^{-8}\; \rm{yr^{-1}}$, we will need to obtain better constrained measurements of $\dot{N}_{\rm TC}$ for a greater number of NSCs. 

\section{Discussion}

\subsection{BH formation in NSCs}

Our analysis finds that NSCs with velocity dispersions higher than $40\;\rm{km\;s^{-1}}$ are X-ray detected at roughly twice the rate of NSCs below that threshold. This is consistent with theories \citep{Miller:2012lr, 2017MNRAS.467.4180S} that suggest NSCs with high velocity dispersions can form and grow massive BHs. 

This is not the only possible interpretation of our results. One alternative scenario is that the NSCs with higher velocity dispersions had pre-existing massive BHs and the NSC formed around the BH. One implication of this scenario is that these NSCs have higher velocity dispersions due to the presence of the BH. To further explore this scenario, we turn to results from \cite{2013ApJ...763...62A}, which models the formation of NSCs through star cluster inspiral in galaxies with/without a pre-existing massive BH. They find that if a massive BH is already present, the resulting NSC has a lower density than would form in a galaxy without a massive BH. In the presence of a pre-existing BH, the radius of the NSC is set by the tidal field of the BH. This could explain why NSCs are not typically found in galaxies with stellar masses greater than $\sim10^{10}\;M_{\odot}$; the tidal field of a sufficently large BH will prevent a NSC from forming\footnote{Other factors, such as the increasingly long dynamical friction inspiral time \citep{2013ApJ...763...62A} and core scouring in multiple mergers \citep{2014ApJ...782...39T}, also likely play a role in suppressing high-mass NSC formation.}. 

\begin{figure}
    \centering
    \includegraphics[width=0.47\textwidth]{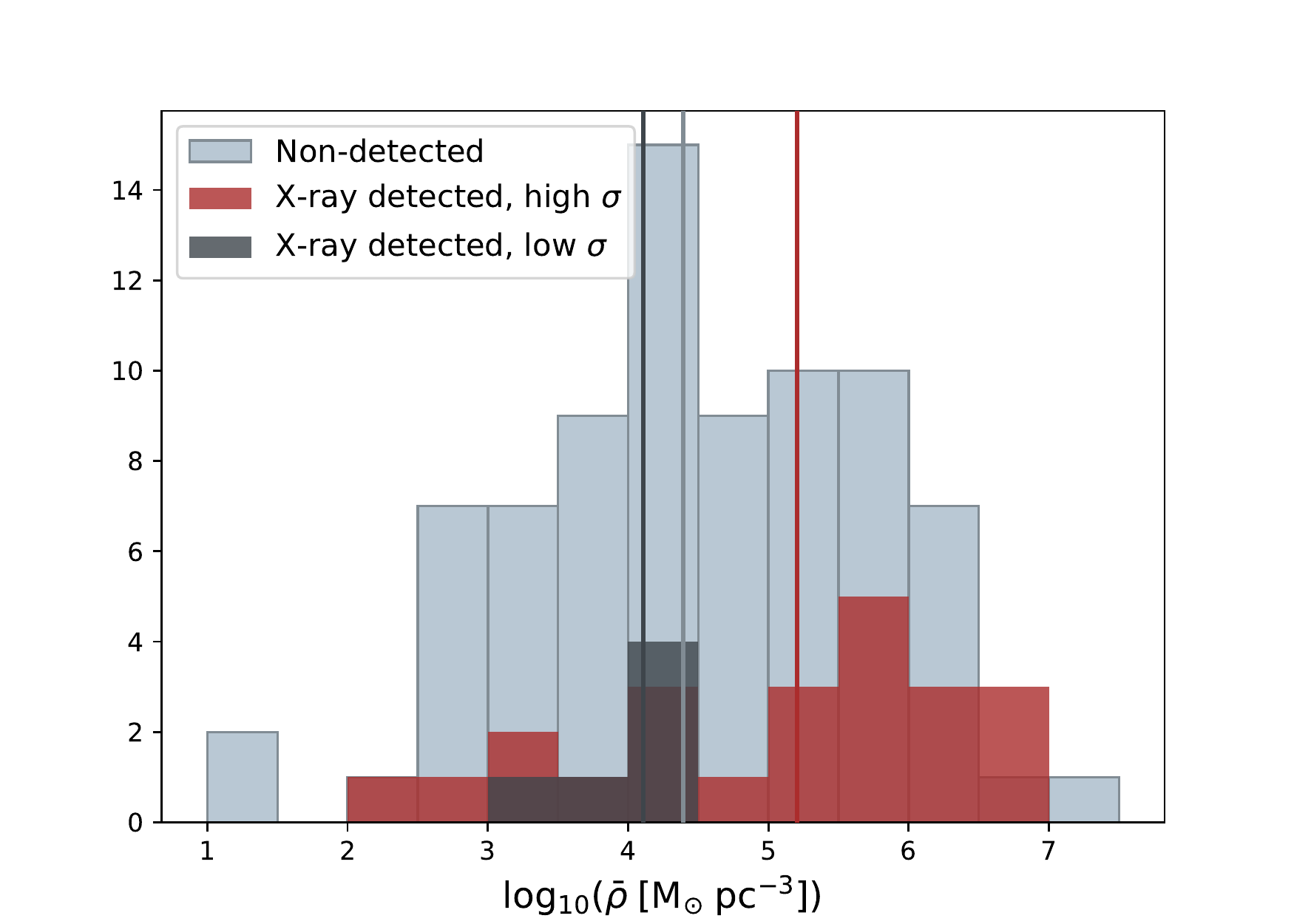}
    \caption{Mean density distributions for NSCs with/without X-ray detections. The mean density is defined as $\bar{\rho} = 3M_{\rm tot} / (4\pi r_{\rm h}^{3})$. The sample of non-detected NSCs is shown in light gray. X-ray detected NSCs with $\sigma>40\rm{km\; s^{-1}}$ are shown in red. X-ray detected NSCs with $\sigma<40\rm{km\; s^{-1}}$ are shown in dark gray. The low-$\sigma$ X-ray detected NSCs have on average lower densities than their high-$\sigma$ counterparts. }
    \label{fig:density}
\end{figure}

If the high-$\sigma$, X-ray detected NSCs formed around pre-existing BHs, they should have systematically lower densities. In Figure~\ref{fig:density} we plot the distributions of mean density for NSCs with and without X-ray detections. For those with X-ray detections, we divide the sample into NSCs with $\sigma$ above and below the 40$\rm{km\;s^{-1}}$ threshold. We find that the X-ray detected, high-$\sigma$ NSCs tend towards \textit{higher} densities, in contrast to expectations for collisionless NSC formation around a pre-existing BH. Interestingly, the X-ray detected NSCs with low $\sigma$ -- those that do not meet the theoretical requirements for forming a massive BH -- skew towards lower densities. Their lower densities are consistent with this population having formed around pre-existing BHs rather than forming a BH later on.

Another interpretation for our findings is that high density/high velocity dispersion NSCs are more efficient at fueling massive BHs, leading to higher accretion rates and thus higher X-ray luminosities. \cite{2015ApJ...803...81N} simulates massive BH growth in NSCs during galaxy mergers. They find that NSCs can enhance BH fueling in this scenario, provided that the NSC meets the condition of $\sigma_{\rm NSC} > v^{2} + c_{\rm s}^{2}$, where $v$ is the relative velocity of the NSC through the gas, and $c_{\rm s}$ is the sound speed. However, this pertains only to systems undergoing mergers; simulations of the impact of NSCs on BH fueling in non-merging systems are needed. 

Results from dynamical studies of nearby nucleated galaxies lend support to the scenario in which massive BHs can form and grow in high velocity-dispersion systems. M33 is one of the most nearby, well studied nuclei; HST kinematics show a NSC velocity dispersion of $24\;\rm{km\;s^{-1}}$ and no detected BH (at least to an upper limit of $1500\;M_{\odot}$; \citealt{2001AJ....122.2469G, 2001Sci...293.1116M}). \cite{2018ApJ...858..118N} carries out dynamical modeling of the nuclear star clusters in M32 (see also \citealt{2010ApJ...725..670S}), NGC 205, NGC 5102, and NGC 5206. The three clusters that have velocity dispersions $>40 \; \rm{km \; s^{-1}}$ -- M32, NGC 5102, and NGC 5206-- are those that are found to have evidence for central black holes. Nearby galaxies with X-ray detections presented here would be good targets for follow-up with high-resolution dynamical modeling.  

These results also have important implications for our understanding of massive BH formation. If NSCs do facilitate the formation of massive BHs at relatively low redshift, this complicates efforts to use the present-day occupation fraction to constrain models of BH formation (see \citealt{2020ARA&A..58..257G}). If massive BH seeds form only at high-redshift, the low-mass end of the occupation fraction is expected to be sensitive to high-redshift BH seed formation models. However, this effect could be washed out if NSCs (which reside in low-mass galaxies) form massive BHs at later epochs.  

\subsection{BH growth in low-mass galaxies}

The most massive BHs are expected to be assembled by $z\sim2$, with progressively lower mass BHs undergoing more of their growth at lower redshifts. This is known as ``downsizing" in BH accretion \citep{2004MNRAS.351..169M, 2008ApJ...676...33D}. In this section, we carry out a simple analysis to explore possible growth histories for the accreting BHs in the sample of nucleated galaxies presented here\footnote{We note that \cite{2014MNRAS.441.3570G} do exclude bright AGN from their initial sample of late-type galaxies based on their AGN class indicator in HyperLeda \citep{2003A&A...412...45P}. This primarily would have removed bright AGN in massive galaxies unlikely to have NSCs. Indeed, NGC 4395 -- one of the brightest and well-known low-mass AGN -- is still part of the sample.}.  

\begin{figure}
    \centering
    \includegraphics[width=0.4\textwidth]{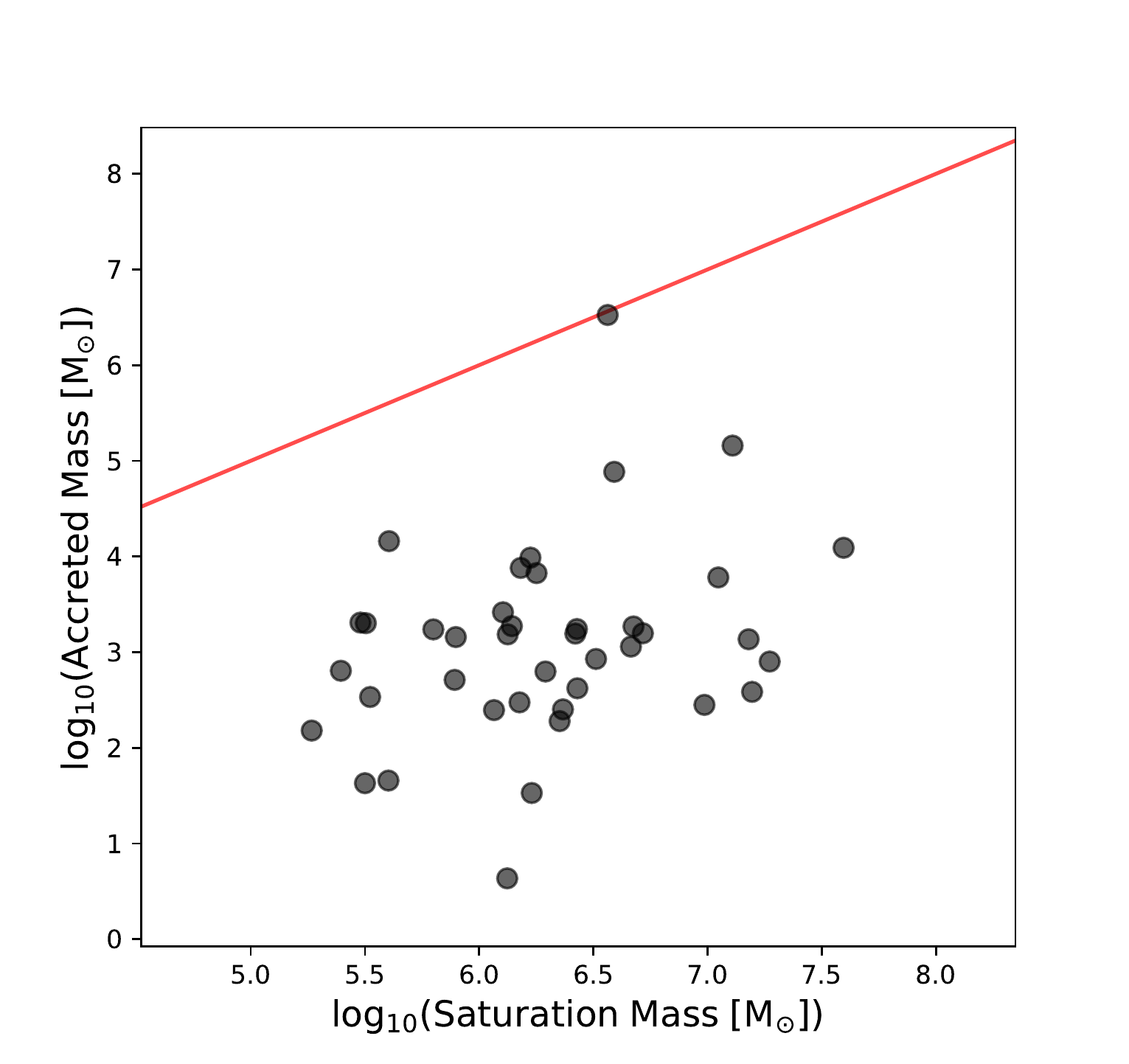}
    \caption{Accreted mass versus saturation mass for each X-ray detected galaxy. The accreted mass is estimated by assuming the present X-ray luminosity can be treated as an average over the last 10 Gyr (see Section 5.2). The saturation mass reflects the mass that a BH in a NSC will grow to through tidal capture and/or tidal disruption processes (equation 37 in \citealt{2017MNRAS.467.4180S}). The red line reflects a saturation mass equal to the accreted mass. }
    \label{fig:bhmass}
\end{figure}

In order to estimate the mass accreted over time, we assume the current 2-10 keV X-ray luminosity for each galaxy is representative and use it as an average X-ray luminosity for that BH over the last 10 Gyr. We assume a bolometric correction of 10 \citep{2004MNRAS.351..169M} and a radiative efficiency of 10\%. Using this analysis, we find accreted masses spanning a large range ($\sim10-10^{6}\;M_{\odot}$), with a mean accreted mass of $\sim10^3\; M_{\odot}$. The mean value is several orders of magnitude below the saturation mass, an order-of-magnitude estimate for BHs grown primarily through tidal capture and/or tidal disruption in NSCs \citep{2017MNRAS.467.4180S}. In Figure~\ref{fig:bhmass}, we show the estimated accreted mass versus the saturation mass for each of the X-ray detected systems. 

This suggests that -- if the current X-ray luminosities can generally be used to compute an average mass accretion rate -- these BHs have not grown much through gas accretion. There are several ways to interpret this result. One possibility is that these BHs were more active in the past and underwent most of their growth at higher redshift. Given the general trend of downsizing in BH accretion, this seems unlikely. Another possibility is that the bolometric correction and/or radiative efficiency are higher for low-mass AGN. However, the accretion properties would have to be substantially different to make up for the several orders of magnitude difference between the estimated accreted mass and saturation mass. A third possibility is that BHs in low-mass galaxies grow primarily through intermittent phases of short, highly super-Eddington accretion due to tidal capture and/or tidal disruption events. While there remain open questions surrounding the fraction of mass accreted in both tidal captures \citep{2018MNRAS.478.4030G} and in tidal disruption events \citep{2016MNRAS.461..948M, 2020MNRAS.495.1374B}, we consider this the most likely scenario. This is consistent with recent work demonstrating that the duty cycles of TDE-powered AGN are consistent with the observed fractions of AGN in present-day dwarf galaxies \citep{2019MNRAS.483.1957Z}.

\section{Summary}

We analyze \textit{Chandra X-ray Observatory} imaging for 108 nearby galaxies with NSCs. Of these, 29 have X-ray emission that is likely to be due to an accreting massive BH. We then study the properties of NSCs with and without evidence for massive BHs. Our conclusions can be summarized as follows:
\begin{enumerate}
    \item NSCs with mean velocity dispersions $>40\; \rm{km\;s^{-1}}$ are X-ray detected at roughly twice the rate of those below this threshold. 
    \item Tentatively, NSCs with TC rates high enough to form an intermediate-mass BH in less than a Hubble time are X-ray detected at a higher rate than those below this limit. 
    \item These results are consistent with a scenario in which sufficiently dense, high velocity dispersion NSCs can form an intermediate-mass BH through dynamical processes.
    \item Better constraints on the active fraction as a function of velocity dispersion and TC rate could be obtained by improved morphological modeling of the NSCs (i.e., finer sampling of the concentration parameters) and additional X-ray observations.
\end{enumerate}

\begin{acknowledgements}

The authors thank Decker French for insightful comments which have improved this manuscript. We thank the anonymous referee for their careful reading of the manuscript and for their suggestions. 

Support for this work was provided by the National Aeronautics and Space Administration through Chandra Award Number GO9-20102X issued by the Chandra X-ray Center, which is operated by the Smithsonian Astrophysical Observatory for and on behalf of the National Aeronautics Space Administration under contract NAS8-03060. The scientific results reported in this article are based on observations made by the Chandra X-ray Observatory through GO-20700424 and data obtained from the Chandra Data Archive. This research has made use of software provided by the Chandra X-ray Center (CXC) in the application packages CIAO, ChIPS, and Sherpa.
\end{acknowledgements}

\bibliographystyle{aasjournal}

\begin{thebibliography}{}
\expandafter\ifx\csname natexlab\endcsname\relax\def\natexlab#1{#1}\fi
\providecommand{\url}[1]{\href{#1}{#1}}
\providecommand{\dodoi}[1]{doi:~\href{http://doi.org/#1}{\nolinkurl{#1}}}
\providecommand{\doeprint}[1]{\href{http://ascl.net/#1}{\nolinkurl{http://ascl.net/#1}}}
\providecommand{\doarXiv}[1]{\href{https://arxiv.org/abs/#1}{\nolinkurl{https://arxiv.org/abs/#1}}}

\bibitem[{{Agarwal} \& {Milosavljevi{\'c}}(2011)}]{Agarwal:2011fk}
{Agarwal}, M., \& {Milosavljevi{\'c}}, M. 2011, \apj, 729, 35,
  \dodoi{10.1088/0004-637X/729/1/35}

\bibitem[{{Antonini}(2013)}]{2013ApJ...763...62A}
{Antonini}, F. 2013, \apj, 763, 62, \dodoi{10.1088/0004-637X/763/1/62}

\bibitem[{{Antonini} {et~al.}(2012){Antonini}, {Capuzzo-Dolcetta},
  {Mastrobuono-Battisti}, \& {Merritt}}]{Antonini:2012uq}
{Antonini}, F., {Capuzzo-Dolcetta}, R., {Mastrobuono-Battisti}, A., \&
  {Merritt}, D. 2012, \apj, 750, 111, \dodoi{10.1088/0004-637X/750/2/111}

\bibitem[{{Antonini} {et~al.}(2019){Antonini}, {Gieles}, \&
  {Gualandris}}]{2019MNRAS.486.5008A}
{Antonini}, F., {Gieles}, M., \& {Gualandris}, A. 2019, \mnras, 486, 5008,
  \dodoi{10.1093/mnras/stz1149}

\bibitem[{{Begelman} {et~al.}(2006){Begelman}, {Volonteri}, \&
  {Rees}}]{2006MNRAS.370..289B}
{Begelman}, M.~C., {Volonteri}, M., \& {Rees}, M.~J. 2006, \mnras, 370, 289,
  \dodoi{10.1111/j.1365-2966.2006.10467.x}

\bibitem[{{Birchall} {et~al.}(2020){Birchall}, {Watson}, \&
  {Aird}}]{2020arXiv200103135B}
{Birchall}, K.~L., {Watson}, M.~G., \& {Aird}, J. 2020, arXiv e-prints,
  arXiv:2001.03135.
\newblock \doarXiv{2001.03135}

\bibitem[{{B{\"o}ker} {et~al.}(2002){B{\"o}ker}, {Laine}, {van der Marel},
  {Sarzi}, {Rix}, {Ho}, \& {Shields}}]{2002AJ....123.1389B}
{B{\"o}ker}, T., {Laine}, S., {van der Marel}, R.~P., {et~al.} 2002, \aj, 123,
  1389, \dodoi{10.1086/339025}

\bibitem[{{B{\"o}ker} {et~al.}(2004){B{\"o}ker}, {Sarzi}, {McLaughlin}, {van
  der Marel}, {Rix}, {Ho}, \& {Shields}}]{2004AJ....127..105B}
{B{\"o}ker}, T., {Sarzi}, M., {McLaughlin}, D.~E., {et~al.} 2004, \aj, 127,
  105, \dodoi{10.1086/380231}

\bibitem[{{Bonnerot} \& {Lu}(2020)}]{2020MNRAS.495.1374B}
{Bonnerot}, C., \& {Lu}, W. 2020, \mnras, 495, 1374,
  \dodoi{10.1093/mnras/staa1246}

\bibitem[{{Capuzzo-Dolcetta} \& {Miocchi}(2008)}]{2008ApJ...681.1136C}
{Capuzzo-Dolcetta}, R., \& {Miocchi}, P. 2008, ApJ, 681, 1136,
  \dodoi{10.1086/588017}

\bibitem[{{Carson} {et~al.}(2015){Carson}, {Barth}, {Seth}, {den Brok},
  {Cappellari}, {Greene}, {Ho}, \& {Neumayer}}]{2015AJ....149..170C}
{Carson}, D.~J., {Barth}, A.~J., {Seth}, A.~C., {et~al.} 2015, \aj, 149, 170,
  \dodoi{10.1088/0004-6256/149/5/170}

\bibitem[{{Clark}(1975)}]{1975ApJ...199L.143C}
{Clark}, G.~W. 1975, \apjl, 199, L143, \dodoi{10.1086/181869}

\bibitem[{{C{\^o}t{\'e}} {et~al.}(2006){C{\^o}t{\'e}}, {Piatek}, {Ferrarese},
  {Jord{\'a}n}, {Merritt}, {Peng}, {Ha{\c s}egan}, {Blakeslee}, {Mei}, {West},
  {Milosavljevi{\'c}}, \& {Tonry}}]{2006ApJS..165...57C}
{C{\^o}t{\'e}}, P., {Piatek}, S., {Ferrarese}, L., {et~al.} 2006, \apjs, 165,
  57, \dodoi{10.1086/504042}

\bibitem[{{Di Matteo} {et~al.}(2008){Di Matteo}, {Colberg}, {Springel},
  {Hernquist}, \& {Sijacki}}]{2008ApJ...676...33D}
{Di Matteo}, T., {Colberg}, J., {Springel}, V., {Hernquist}, L., \& {Sijacki},
  D. 2008, \apj, 676, 33, \dodoi{10.1086/524921}

\bibitem[{{Ebisuzaki} {et~al.}(1991){Ebisuzaki}, {Makino}, \&
  {Okumura}}]{1991Natur.354..212E}
{Ebisuzaki}, T., {Makino}, J., \& {Okumura}, S.~K. 1991, \nat, 354, 212,
  \dodoi{10.1038/354212a0}

\bibitem[{{Fabian} {et~al.}(1975){Fabian}, {Pringle}, \&
  {Rees}}]{1975MNRAS.172P..15F}
{Fabian}, A.~C., {Pringle}, J.~E., \& {Rees}, M.~J. 1975, \mnras, 172, 15,
  \dodoi{10.1093/mnras/172.1.15P}

\bibitem[{{Fahrion} {et~al.}(2021){Fahrion}, {Lyubenova}, {van de Ven},
  {Hilker}, {Leaman}, {Falc{\'o}n-Barroso}, {Bittner}, {Coccato}, {Corsini},
  {Gadotti}, {Iodice}, {McDermid}, {Mart{\'\i}n-Navarro}, {Pinna}, {Poci},
  {Sarzi}, {de Zeeuw}, \& {Zhu}}]{2021arXiv210406412F}
{Fahrion}, K., {Lyubenova}, M., {van de Ven}, G., {et~al.} 2021, arXiv
  e-prints, arXiv:2104.06412.
\newblock \doarXiv{2104.06412}

\bibitem[{Ferrarese {et~al.}(2006{\natexlab{a}})Ferrarese, {C{\^o}t{\'e}},
  {Jord{\'a}n}, Peng, Slakeslee, Piatek, Mei, Merritt, Milosavljevic, Tonry, \&
  West}]{:fs}
Ferrarese, L., {C{\^o}t{\'e}}, P., {Jord{\'a}n}, A., {et~al.}
  2006{\natexlab{a}}, ApJ, 164, 334

\bibitem[{Ferrarese {et~al.}(2006{\natexlab{b}})Ferrarese, {C{\^o}t{\'e}},
  Bonta, Peng, Merritt, Jordan, Blakeslee, Hasegan, Mei, Piatek, Tonry, \&
  West}]{:uw}
Ferrarese, L., {C{\^o}t{\'e}}, P., Bonta, E.~D., {et~al.} 2006{\natexlab{b}},
  ApJ, 644, L21

\bibitem[{{Foord} {et~al.}(2017){Foord}, {Gallo}, {Hodges-Kluck}, {Miller},
  {Baldassare}, {G{\"u}ltekin}, \& {Gnedin}}]{2017ApJ...841...51F}
{Foord}, A., {Gallo}, E., {Hodges-Kluck}, E., {et~al.} 2017, \apj, 841, 51,
  \dodoi{10.3847/1538-4357/aa6d63}

\bibitem[{{Fragione} {et~al.}(2021){Fragione}, {Kocsis}, {Rasio}, \&
  {Silk}}]{2021arXiv210704639F}
{Fragione}, G., {Kocsis}, B., {Rasio}, F.~A., \& {Silk}, J. 2021, arXiv
  e-prints, arXiv:2107.04639.
\newblock \doarXiv{2107.04639}

\bibitem[{{Fragione} \& {Silk}(2020)}]{2020MNRAS.498.4591F}
{Fragione}, G., \& {Silk}, J. 2020, \mnras, 498, 4591,
  \dodoi{10.1093/mnras/staa2629}

\bibitem[{{Gebhardt} {et~al.}(2001){Gebhardt}, {Lauer}, {Kormendy}, {Pinkney},
  {Bower}, {Green}, {Gull}, {Hutchings}, {Kaiser}, {Nelson}, {Richstone}, \&
  {Weistrop}}]{2001AJ....122.2469G}
{Gebhardt}, K., {Lauer}, T.~R., {Kormendy}, J., {et~al.} 2001, \aj, 122, 2469,
  \dodoi{10.1086/323481}

\bibitem[{{Gehrels}(1986)}]{Gehrels:1986kx}
{Gehrels}, N. 1986, \apj, 303, 336, \dodoi{10.1086/164079}

\bibitem[{{Generozov} {et~al.}(2018){Generozov}, {Stone}, {Metzger}, \&
  {Ostriker}}]{2018MNRAS.478.4030G}
{Generozov}, A., {Stone}, N.~C., {Metzger}, B.~D., \& {Ostriker}, J.~P. 2018,
  \mnras, 478, 4030, \dodoi{10.1093/mnras/sty1262}

\bibitem[{{Georgiev} \& {B{\"o}ker}(2014)}]{2014MNRAS.441.3570G}
{Georgiev}, I.~Y., \& {B{\"o}ker}, T. 2014, \mnras, 441, 3570,
  \dodoi{10.1093/mnras/stu797}

\bibitem[{{Georgiev} {et~al.}(2016){Georgiev}, {B{\"o}ker}, {Leigh},
  {L{\"u}tzgendorf}, \& {Neumayer}}]{2016MNRAS.457.2122G}
{Georgiev}, I.~Y., {B{\"o}ker}, T., {Leigh}, N., {L{\"u}tzgendorf}, N., \&
  {Neumayer}, N. 2016, \mnras, 457, 2122, \dodoi{10.1093/mnras/stw093}

\bibitem[{Gilfanov(2004)}]{:ut}
Gilfanov, M. 2004, MNRAS, 349, 146

\bibitem[{{Gnedin} {et~al.}(2014){Gnedin}, {Ostriker}, \&
  {Tremaine}}]{2014ApJ...785...71G}
{Gnedin}, O.~Y., {Ostriker}, J.~P., \& {Tremaine}, S. 2014, \apj, 785, 71,
  \dodoi{10.1088/0004-637X/785/1/71}

\bibitem[{{Gonz{\'a}lez} {et~al.}(2021){Gonz{\'a}lez}, {Kremer}, {Chatterjee},
  {Fragione}, {Rodriguez}, {Weatherford}, {Ye}, \&
  {Rasio}}]{2021ApJ...908L..29G}
{Gonz{\'a}lez}, E., {Kremer}, K., {Chatterjee}, S., {et~al.} 2021, \apjl, 908,
  L29, \dodoi{10.3847/2041-8213/abdf5b}

\bibitem[{{Greene} {et~al.}(2020){Greene}, {Strader}, \&
  {Ho}}]{2020ARA&A..58..257G}
{Greene}, J.~E., {Strader}, J., \& {Ho}, L.~C. 2020, \araa, 58, 257,
  \dodoi{10.1146/annurev-astro-032620-021835}

\bibitem[{{Grimm} {et~al.}(2003){Grimm}, {Gilfanov}, \&
  {Sunyaev}}]{2003MNRAS.339..793G}
{Grimm}, H.-J., {Gilfanov}, M., \& {Sunyaev}, R. 2003, \mnras, 339, 793,
  \dodoi{10.1046/j.1365-8711.2003.06224.x}

\bibitem[{{G{\"u}rkan} {et~al.}(2004){G{\"u}rkan}, {Freitag}, \&
  {Rasio}}]{2004ApJ...604..632G}
{G{\"u}rkan}, M.~A., {Freitag}, M., \& {Rasio}, F.~A. 2004, \apj, 604, 632,
  \dodoi{10.1086/381968}

\bibitem[{{Hailey} {et~al.}(2018){Hailey}, {Mori}, {Bauer}, {Berkowitz},
  {Hong}, \& {Hord}}]{2018Natur.556...70H}
{Hailey}, C.~J., {Mori}, K., {Bauer}, F.~E., {et~al.} 2018, \nat, 556, 70,
  \dodoi{10.1038/nature25029}

\bibitem[{{Haiman} \& {Loeb}(2001)}]{2001ApJ...552..459H}
{Haiman}, Z., \& {Loeb}, A. 2001, \apj, 552, 459, \dodoi{10.1086/320586}

\bibitem[{{Hannah} {et~al.}(2021){Hannah}, {Seth}, {Nguyen}, {Dumont},
  {Kacharov}, {Neumayer}, \& {den Brok}}]{2021arXiv210912251H}
{Hannah}, C.~H., {Seth}, A.~C., {Nguyen}, D.~D., {et~al.} 2021, arXiv e-prints,
  arXiv:2109.12251.
\newblock \doarXiv{2109.12251}

\bibitem[{{Hoyer} {et~al.}(2021){Hoyer}, {Neumayer}, {Georgiev}, {Seth}, \&
  {Greene}}]{2021arXiv210705313H}
{Hoyer}, N., {Neumayer}, N., {Georgiev}, I.~Y., {Seth}, A.~C., \& {Greene},
  J.~E. 2021, arXiv e-prints, arXiv:2107.05313.
\newblock \doarXiv{2107.05313}

\bibitem[{{Inayoshi} {et~al.}(2020){Inayoshi}, {Visbal}, \&
  {Haiman}}]{2020ARA&A..58...27I}
{Inayoshi}, K., {Visbal}, E., \& {Haiman}, Z. 2020, \araa, 58, 27,
  \dodoi{10.1146/annurev-astro-120419-014455}

\bibitem[{{Ivanova} {et~al.}(2008){Ivanova}, {Heinke}, {Rasio}, {Belczynski},
  \& {Fregeau}}]{2008MNRAS.386..553I}
{Ivanova}, N., {Heinke}, C.~O., {Rasio}, F.~A., {Belczynski}, K., \& {Fregeau},
  J.~M. 2008, \mnras, 386, 553, \dodoi{10.1111/j.1365-2966.2008.13064.x}

\bibitem[{{Kennicutt} \& {Evans}(2012)}]{2012ARAA..50..531K}
{Kennicutt}, R.~C., \& {Evans}, N.~J. 2012, \araa, 50, 531,
  \dodoi{10.1146/annurev-astro-081811-125610}

\bibitem[{{King}(1966)}]{King:1966wd}
{King}, I.~R. 1966, \aj, 71, 64, \dodoi{10.1086/109857}

\bibitem[{{Kormendy} \& {Ho}(2013)}]{Kormendy:2013ve}
{Kormendy}, J., \& {Ho}, L.~C. 2013, \araa, 51, 511,
  \dodoi{10.1146/annurev-astro-082708-101811}

\bibitem[{{Kremer} {et~al.}(2020){Kremer}, {Spera}, {Becker}, {Chatterjee}, {Di
  Carlo}, {Fragione}, {Rodriguez}, {Ye}, \& {Rasio}}]{2020ApJ...903...45K}
{Kremer}, K., {Spera}, M., {Becker}, D., {et~al.} 2020, \apj, 903, 45,
  \dodoi{10.3847/1538-4357/abb945}

\bibitem[{{Latif} {et~al.}(2013){Latif}, {Schleicher}, {Schmidt}, \&
  {Niemeyer}}]{2013MNRAS.436.2989L}
{Latif}, M.~A., {Schleicher}, D.~R.~G., {Schmidt}, W., \& {Niemeyer}, J.~C.
  2013, \mnras, 436, 2989, \dodoi{10.1093/mnras/stt1786}

\bibitem[{{Lee} \& {Ostriker}(1986)}]{1986ApJ...310..176L}
{Lee}, H.~M., \& {Ostriker}, J.~P. 1986, \apj, 310, 176, \dodoi{10.1086/164674}

\bibitem[{{Lee} {et~al.}(2019){Lee}, {Gallo}, {Hodges-Kluck}, {Cot{\'e}},
  {Ferrarese}, {Miller}, {Baldassare}, {Plotkin}, \&
  {Treu}}]{2019ApJ...874...77L}
{Lee}, N., {Gallo}, E., {Hodges-Kluck}, E., {et~al.} 2019, \apj, 874, 77,
  \dodoi{10.3847/1538-4357/ab05cd}

\bibitem[{{Lehmer} {et~al.}(2019){Lehmer}, {Eufrasio}, {Tzanavaris},
  {Basu-Zych}, {Fragos}, {Prestwich}, {Yukita}, {Zezas}, {Hornschemeier}, \&
  {Ptak}}]{2019ApJS..243....3L}
{Lehmer}, B.~D., {Eufrasio}, R.~T., {Tzanavaris}, P., {et~al.} 2019, \apjs,
  243, 3, \dodoi{10.3847/1538-4365/ab22a8}

\bibitem[{{Leigh} {et~al.}(2012){Leigh}, {B{\"o}ker}, \&
  {Knigge}}]{2012MNRAS.424.2130L}
{Leigh}, N., {B{\"o}ker}, T., \& {Knigge}, C. 2012, \mnras, 424, 2130,
  \dodoi{10.1111/j.1365-2966.2012.21365.x}

\bibitem[{{Loeb} \& {Rasio}(1994)}]{1994ApJ...432...52L}
{Loeb}, A., \& {Rasio}, F.~A. 1994, \apj, 432, 52, \dodoi{10.1086/174548}

\bibitem[{{Lotz} {et~al.}(2001){Lotz}, {Telford}, {Ferguson}, {Miller},
  {Stiavelli}, \& {Mack}}]{Lotz:2001zr}
{Lotz}, J.~M., {Telford}, R., {Ferguson}, H.~C., {et~al.} 2001, \apj, 552, 572,
  \dodoi{10.1086/320545}

\bibitem[{{Madau} \& {Rees}(2001)}]{2001ApJ...551L..27M}
{Madau}, P., \& {Rees}, M.~J. 2001, \apjl, 551, L27, \dodoi{10.1086/319848}

\bibitem[{{Marconi} {et~al.}(2004){Marconi}, {Risaliti}, {Gilli}, {Hunt},
  {Maiolino}, \& {Salvati}}]{2004MNRAS.351..169M}
{Marconi}, A., {Risaliti}, G., {Gilli}, R., {et~al.} 2004, \mnras, 351, 169,
  \dodoi{10.1111/j.1365-2966.2004.07765.x}

\bibitem[{{Martin} {et~al.}(2005){Martin}, {Fanson}, {Schiminovich},
  {Morrissey}, {Friedman}, {Barlow}, {Conrow}, {Grange}, {Jelinsky},
  {Milliard}, {Siegmund}, {Bianchi}, {Byun}, {Donas}, {Forster}, {Heckman},
  {Lee}, {Madore}, {Malina}, {Neff}, {Rich}, {Small}, {Surber}, {Szalay},
  {Welsh}, \& {Wyder}}]{2005ApJ...619L...1M}
{Martin}, D.~C., {Fanson}, J., {Schiminovich}, D., {et~al.} 2005, \apjl, 619,
  L1, \dodoi{10.1086/426387}

\bibitem[{{Mayer} {et~al.}(2010){Mayer}, {Kazantzidis}, {Escala}, \&
  {Callegari}}]{2010Natur.466.1082M}
{Mayer}, L., {Kazantzidis}, S., {Escala}, A., \& {Callegari}, S. 2010, \nat,
  466, 1082, \dodoi{10.1038/nature09294}

\bibitem[{{Merritt} {et~al.}(2001){Merritt}, {Ferrarese}, \&
  {Joseph}}]{2001Sci...293.1116M}
{Merritt}, D., {Ferrarese}, L., \& {Joseph}, C.~L. 2001, Science, 293, 1116,
  \dodoi{10.1126/science.1063896}

\bibitem[{{Metzger} \& {Stone}(2016)}]{2016MNRAS.461..948M}
{Metzger}, B.~D., \& {Stone}, N.~C. 2016, \mnras, 461, 948,
  \dodoi{10.1093/mnras/stw1394}

\bibitem[{{Miller} \& {Davies}(2012)}]{Miller:2012lr}
{Miller}, M.~C., \& {Davies}, M.~B. 2012, \apj, 755, 81,
  \dodoi{10.1088/0004-637X/755/1/81}

\bibitem[{{Mineo} {et~al.}(2012){Mineo}, {Gilfanov}, \&
  {Sunyaev}}]{2012MNRAS.419.2095M}
{Mineo}, S., {Gilfanov}, M., \& {Sunyaev}, R. 2012, \mnras, 419, 2095,
  \dodoi{10.1111/j.1365-2966.2011.19862.x}

\bibitem[{{Muno} {et~al.}(2005){Muno}, {Pfahl}, {Baganoff}, {Brandt}, {Ghez},
  {Lu}, \& {Morris}}]{2005ApJ...622L.113M}
{Muno}, M.~P., {Pfahl}, E., {Baganoff}, F.~K., {et~al.} 2005, \apjl, 622, L113,
  \dodoi{10.1086/429721}

\bibitem[{{Naiman} {et~al.}(2015){Naiman}, {Ramirez-Ruiz}, {Debuhr}, \&
  {Ma}}]{2015ApJ...803...81N}
{Naiman}, J.~P., {Ramirez-Ruiz}, E., {Debuhr}, J., \& {Ma}, C.-P. 2015, \apj,
  803, 81, \dodoi{10.1088/0004-637X/803/2/81}

\bibitem[{{Natarajan}(2021)}]{2021MNRAS.501.1413N}
{Natarajan}, P. 2021, \mnras, 501, 1413, \dodoi{10.1093/mnras/staa3724}

\bibitem[{{Neumayer} {et~al.}(2020){Neumayer}, {Seth}, \&
  {B{\"o}ker}}]{2020A&ARv..28....4N}
{Neumayer}, N., {Seth}, A., \& {B{\"o}ker}, T. 2020, \aapr, 28, 4,
  \dodoi{10.1007/s00159-020-00125-0}

\bibitem[{{Nguyen} {et~al.}(2018){Nguyen}, {Seth}, {Neumayer}, {Kamann},
  {Voggel}, {Cappellari}, {Picotti}, {Nguyen}, {B{\"o}ker}, {Debattista},
  {Caldwell}, {McDermid}, {Bastian}, {Ahn}, \&
  {Pechetti}}]{2018ApJ...858..118N}
{Nguyen}, D.~D., {Seth}, A.~C., {Neumayer}, N., {et~al.} 2018, \apj, 858, 118,
  \dodoi{10.3847/1538-4357/aabe28}

\bibitem[{{Nguyen} {et~al.}(2021){Nguyen}, {Bureau}, {Thater}, {Nyland}, {den
  Brok}, {Cappellari}, {Davis}, {Greene}, {Neumayer}, {Imanishi}, {Izumi},
  {Kawamuro}, {Baba}, {Nguyen}, {Iguchi}, {Tsukui}, {Lam N.}, \&
  {Ho}}]{2021arXiv211008476N}
{Nguyen}, D.~D., {Bureau}, M., {Thater}, S., {et~al.} 2021, arXiv e-prints,
  arXiv:2110.08476.
\newblock \doarXiv{2110.08476}

\bibitem[{{Paturel} {et~al.}(2003){Paturel}, {Petit}, {Prugniel}, {Theureau},
  {Rousseau}, {Brouty}, {Dubois}, \& {Cambr{\'e}sy}}]{2003A&A...412...45P}
{Paturel}, G., {Petit}, C., {Prugniel}, P., {et~al.} 2003, \aap, 412, 45,
  \dodoi{10.1051/0004-6361:20031411}

\bibitem[{{Portegies Zwart} {et~al.}(2004){Portegies Zwart}, {Baumgardt},
  {Hut}, {Makino}, \& {McMillan}}]{2004Natur.428..724P}
{Portegies Zwart}, S.~F., {Baumgardt}, H., {Hut}, P., {Makino}, J., \&
  {McMillan}, S.~L.~W. 2004, \nat, 428, 724, \dodoi{10.1038/nature02448}

\bibitem[{{Portegies Zwart} \& {McMillan}(2002)}]{2002ApJ...576..899P}
{Portegies Zwart}, S.~F., \& {McMillan}, S. L.~W. 2002, \apj, 576, 899,
  \dodoi{10.1086/341798}

\bibitem[{{Pouliasis} {et~al.}(2019){Pouliasis}, {Georgantopoulos}, {Bonanos},
  {Yang}, {Sokolovsky}, {Hatzidimitriou}, {Mountrichas}, {Gavras},
  {Charmandaris}, {Bellas-Velidis}, {Spetsieri}, \&
  {Tsinganos}}]{2019MNRAS.487.4285P}
{Pouliasis}, E., {Georgantopoulos}, I., {Bonanos}, A.~Z., {et~al.} 2019,
  \mnras, 487, 4285, \dodoi{10.1093/mnras/stz1483}

\bibitem[{{Quinlan}(1996)}]{1996NewA....1...35Q}
{Quinlan}, G.~D. 1996, \na, 1, 35, \dodoi{10.1016/S1384-1076(96)00003-6}

\bibitem[{{S{\'a}nchez-Janssen} {et~al.}(2019){S{\'a}nchez-Janssen},
  {C{\^o}t{\'e}}, {Ferrarese}, {Peng}, {Roediger}, {Blakeslee}, {Emsellem},
  {Puzia}, {Spengler}, {Taylor}, {{\'A}lamo-Mart{\'\i}nez}, {Boselli},
  {Cantiello}, {Cuillandre}, {Duc}, {Durrell}, {Gwyn}, {MacArthur},
  {Lan{\c{c}}on}, {Lim}, {Liu}, {Mei}, {Miller}, {Mu{\~n}oz}, {Mihos},
  {Paudel}, {Powalka}, \& {Toloba}}]{2019ApJ...878...18S}
{S{\'a}nchez-Janssen}, R., {C{\^o}t{\'e}}, P., {Ferrarese}, L., {et~al.} 2019,
  \apj, 878, 18, \dodoi{10.3847/1538-4357/aaf4fd}

\bibitem[{{Sch{\"o}del} {et~al.}(2009){Sch{\"o}del}, {Merritt}, \&
  {Eckart}}]{2009A&A...502...91S}
{Sch{\"o}del}, R., {Merritt}, D., \& {Eckart}, A. 2009, \aap, 502, 91,
  \dodoi{10.1051/0004-6361/200810922}

\bibitem[{{Seth} {et~al.}(2008){Seth}, {Ag{\"u}eros}, {Lee}, \&
  {Basu-Zych}}]{2008ApJ...678..116S}
{Seth}, A., {Ag{\"u}eros}, M., {Lee}, D., \& {Basu-Zych}, A. 2008, \apj, 678,
  116, \dodoi{10.1086/528955}

\bibitem[{{Seth}(2010)}]{2010ApJ...725..670S}
{Seth}, A.~C. 2010, \apj, 725, 670, \dodoi{10.1088/0004-637X/725/1/670}

\bibitem[{{Seth} {et~al.}(2006){Seth}, {Dalcanton}, {Hodge}, \&
  {Debattista}}]{2006AJ....132.2539S}
{Seth}, A.~C., {Dalcanton}, J.~J., {Hodge}, P.~W., \& {Debattista}, V.~P. 2006,
  \aj, 132, 2539, \dodoi{10.1086/508994}

\bibitem[{Sivakoff {et~al.}(2007)Sivakoff, {Jord{\'a}n}, Sarazin, Blakeslee,
  Cote, Ferrarese, Juett, Mei, \& Peng}]{:yf}
Sivakoff, G., {Jord{\'a}n}, A., Sarazin, C., {et~al.} 2007, ApJ, 660, 1246

\bibitem[{{Stone} {et~al.}(2017){Stone}, {K{\"u}pper}, \&
  {Ostriker}}]{2017MNRAS.467.4180S}
{Stone}, N.~C., {K{\"u}pper}, A.~H.~W., \& {Ostriker}, J.~P. 2017, \mnras, 467,
  4180, \dodoi{10.1093/mnras/stx097}

\bibitem[{{Thomas} {et~al.}(2014){Thomas}, {Saglia}, {Bender}, {Erwin}, \&
  {Fabricius}}]{2014ApJ...782...39T}
{Thomas}, J., {Saglia}, R.~P., {Bender}, R., {Erwin}, P., \& {Fabricius}, M.
  2014, \apj, 782, 39, \dodoi{10.1088/0004-637X/782/1/39}

\bibitem[{{Tremaine} {et~al.}(1975){Tremaine}, {Ostriker}, \&
  {Spitzer}}]{1975ApJ...196..407T}
{Tremaine}, S.~D., {Ostriker}, J.~P., \& {Spitzer}, Jr., L. 1975, \apj, 196,
  407, \dodoi{10.1086/153422}

\bibitem[{Turner {et~al.}(2012)Turner, Cote, Ferrarese, Jordan, Blakeslee, Mei,
  Peng, \& West}]{:fz}
Turner, M., Cote, P., Ferrarese, L., {et~al.} 2012, ApJS

\bibitem[{{Volonteri}(2010)}]{2010A&ARv..18..279V}
{Volonteri}, M. 2010, \aapr, 18, 279, \dodoi{10.1007/s00159-010-0029-x}

\bibitem[{{Walcher} {et~al.}(2006){Walcher}, {B{\"o}ker}, {Charlot}, {Ho},
  {Rix}, {Rossa}, {Shields}, \& {van der Marel}}]{2006ApJ...649..692W}
{Walcher}, C.~J., {B{\"o}ker}, T., {Charlot}, S., {et~al.} 2006, \apj, 649,
  692, \dodoi{10.1086/505166}

\bibitem[{{Whalen} \& {Fryer}(2012)}]{2012ApJ...756L..19W}
{Whalen}, D.~J., \& {Fryer}, C.~L. 2012, \apjl, 756, L19,
  \dodoi{10.1088/2041-8205/756/1/L19}

\bibitem[{{Wright} {et~al.}(2010){Wright}, {Eisenhardt}, {Mainzer}, {Ressler},
  {Cutri}, {Jarrett}, {Kirkpatrick}, {Padgett}, {McMillan}, {Skrutskie},
  {Stanford}, {Cohen}, {Walker}, {Mather}, {Leisawitz}, {Gautier}, {McLean},
  {Benford}, {Lonsdale}, {Blain}, {Mendez}, {Irace}, {Duval}, {Liu}, {Royer},
  {Heinrichsen}, {Howard}, {Shannon}, {Kendall}, {Walsh}, {Larsen}, {Cardon},
  {Schick}, {Schwalm}, {Abid}, {Fabinsky}, {Naes}, \&
  {Tsai}}]{2010AJ....140.1868W}
{Wright}, E.~L., {Eisenhardt}, P. R.~M., {Mainzer}, A.~K., {et~al.} 2010, \aj,
  140, 1868, \dodoi{10.1088/0004-6256/140/6/1868}

\bibitem[{{Yang} {et~al.}(2015){Yang}, {Xie}, {Yuan}, {Zdziarski},
  {Gierli{\'n}ski}, {Ho}, \& {Yu}}]{2015MNRAS.447.1692Y}
{Yang}, Q.-X., {Xie}, F.-G., {Yuan}, F., {et~al.} 2015, \mnras, 447, 1692,
  \dodoi{10.1093/mnras/stu2571}

\bibitem[{{Zubovas}(2019)}]{2019MNRAS.483.1957Z}
{Zubovas}, K. 2019, \mnras, 483, 1957, \dodoi{10.1093/mnras/sty3211}

\end{thebibliography}

\end{document}